\documentclass[preprint,12pt,authoryear]{elsarticle}
\usepackage{graphics,bm,graphicx}
\usepackage{float}
\usepackage{epstopdf}
\usepackage{tikz}
\usepackage{pdfpages}
\usepackage{color,soul}
\usepackage{subfig}
\usepackage{mathtools}
\usepackage{amsmath,mathptmx}
\usepackage{amssymb}
\usepackage{multicol}
\usepackage{booktabs}
\usepackage{tabularx}
\renewcommand{\arraystretch}{1.5}
\makeatletter
\def\ps@pprintTitle{%
}
\usepackage[colorlinks]{hyperref}
\AtBeginDocument{%
	\hypersetup{%
		linkcolor=black,%
		citecolor=black,%
	}%
}%

\begin{document}

\begin{frontmatter}
	
	\title{The role of magnetic waves in tangent cylinder convection}
	\author{Debarshi Majumder}
	\author{Binod Sreenivasan \corref{cor1}}
	\cortext[cor1]{Corresponding author}
	\ead{bsreeni@iisc.ac.in}
	\address{Centre for Earth Sciences, 
		Indian Institute of Science, Bangalore 560012, India}
\begin{abstract}
The secular variation of the geomagnetic field
suggests that there are anticyclonic
polar vortices
in the Earth's core. Under the influence of
a magnetic field, the polar azimuthal
flow is
thought to be produced by one or more coherent
upwellings within the tangent cylinder, offset
from the rotation axis.
In this study, 
convection within the tangent cylinder in rapidly
rotating dynamos is investigated through the 
analysis of forced magnetic waves. 
The first part of the study investigates 
the evolution of an isolated buoyancy disturbance 
in an unstably stratified rotating 
fluid subjected to an axial magnetic field. 
It is shown that the axial flow intensity of 
the slow Magnetic-Archimedean-Coriolis (MAC) 
waves becomes comparable to that of the fast 
MAC waves when $|\omega_M/\omega_C| \sim 0.1$, 
where $\omega_M$ and $\omega_C$ are
the Alfv\'en wave and inertial wave frequencies
respectively. In spherical shell dynamo
simulations, the isolated
upwellings within the tangent cylinder
are shown to originate from
the localized excitation of slow MAC waves in the dipole-dominated
regime. Axial flow measurements in turn
reveal the approximate parity between the slow and fast
wave intensities in this regime, which corresponds to 
the existence of strong polar vortices 
in the Earth's core. 
To obtain the observed
peak azimuthal motions of $0.6$--$0.9^\circ$yr$^{-1}$,
the Rayleigh number in the low-inertia
geodynamo must be $\sim 10^3$ 
times the Rayleigh number
for the onset of nonmagnetic convection. 
However, if the forcing is so
strong as to cause polarity reversals, the field
within the tangent cylinder decays away, and the
convection takes the form of an ensemble of plumes
supported entirely by the fast waves of frequency
$\omega \sim \omega_C$.
The resulting weak polar circulation is comparable
to that obtained in nonmagnetic convection.

		\end{abstract}


\end{frontmatter}

\section{Introduction}
Convection in the Earth's outer core is separated into
two regions by rapid rotation, inside and outside the
tangent cylinder (TC). The TC is an imaginary cylinder
tangent to the inner core boundary (ICB) 
and parallel to
the Earth's rotation axis $z$. It cuts the core--mantle
boundary (CMB) at approximately latitude 70$^\circ$.
Inside the TC, the heat and compositional flux have
a substantial component in the $z$ direction, so fluid
motions are strongly $z$-dependent.
Because ageostrophic motions are needed to transport
heat and light elements
 from the ICB to the CMB inside the TC \citep{jones15}, the
forcing needed to initiate convection inside the TC is
much higher than that outside it.

The convection inside the TC
is thought to be instrumental in the generation 
of anticyclonic polar vortices in the core,  suggested
by observations of secular variation \citep{olson1999polar,hulot2002}.
Nonmagnetic laboratory experiments
that simulate the TC region \citep{aurnou2003expt,aujogue2018}
suggest that
an ensemble of convection plumes extending from the 
ICB to latitudes greater than 70$^\circ$
 would make the poles warmer
than the equator, resulting in an anticyclonic
circulation near the poles. Numerical simulations
of the geodynamo \citep{grl2005,gafd2006}, on the
other hand, show that the structure of convection
inside the TC is often dominated by an 
isolated off-axis plume
that generates a strong anticyclonic flow
near the poles. The fluid inside the
rising plume is systematically warmer than the cold 
descending fluid outside the plume.
The Coriolis force then acts to turn the 
radially outward flow at the top of
the plume into an anticyclonic vortex. 
The intensity of polar vortices in the dynamo
is much greater than that in nonmagnetic
convection \citep{grl2005}, which indicates
that the vortices are magnetically enhanced. 

The tangent cylinder may be approximated by a 
rotating
plane layer in which convection takes 
place under a predominantly $z$ magnetic
field. The onset of magnetoconvection 
in a rotating plane layer occurs either as thin viscously 
controlled columns or large-scale
magnetic rolls \citep{chandra61}. For a field that is either uniform or
of a length scale comparable to the depth of the fluid layer, 
large-scale magnetically
controlled convection sets in at relatively
small Elsasser numbers $\varLambda = O(E^{1/3})$
\citep{zhang95,jones2003}, where $\varLambda$ is the square
of the scaled mean magnetic field and $E$ is the Ekman number that
measures the ratio of viscous to Coriolis forces. 
However, for the spatially inhomogeneous magnetic field in a
dynamo \citep[e.g.][]{schaeff2017}, 
the viscous--magnetic mode transition may not take place
even at $\varLambda =O(1)$ \citep{venka2015}. In a dipole-dominated
dynamo, the field generated outside the TC diffuses into the
TC and gets concentrated on scales comparable to that of
the thin plumes that form at the onset of
convection. The
lateral inhomogeneity of the magnetic flux gives rise to
an instability where convection is entirely confined
to the peak-field region \citep{jfm17a}. This pattern
of convection inside the TC may be understood in
terms of fast and slow Magnetic-Archimedean-Coriolis (MAC)
waves that form in a rapidly rotating B\'enard layer
permeated by a magnetic field. 
The fast waves are linear inertial waves weakly
modified by the magnetic field and buoyancy while the 
slow waves are magnetostrophic waves
produced by the balance
 between the magnetic, Coriolis and buoyancy forces 
\citep{braginsky1967,acheson1973}. The magnetic
flux concentrations within the TC may locally produce
unstable stratification, thereby supporting convection
through slow MAC waves. In regions where the
magnetic flux is weak, the fast MAC waves generated
at the base of the TC
are rendered
ineffectual in transporting heat and light elements
through the neutrally buoyant layer. In this way,
a laterally varying magnetic field can give rise
to an isolated off-axis plume within the TC. If the
buoyant forcing in the dynamo is so strong as to cause the collapse
of the axial dipole field outside the TC,
the field intensity within the TC would be considerably
reduced. Consequently, the convection may take the form of
an ensemble of plumes supported by fast waves, 
akin to that found in the nonmagnetic state. 

The present study builds on earlier work \citep{jfm21} 
that investigated the evolution of an isolated buoyancy 
disturbance in a rapidly
rotating fluid under a uniform axial
magnetic field. Of particular interest is
the regime where the ratio of Alfv\'en wave
to inertial wave frequencies $|\omega_M/\omega_C| \sim 0.1$,
thought to be relevant to dipole-dominated
dynamos. Subsequently, \cite{jfm23} demonstrated 
that the collapse of the axial dipole in rapidly rotating 
dynamos occurs when the slow MAC waves disappear 
under strong forcing. Here, TC convection in dipole-dominated 
and polarity-reversing dynamos is understood 
through the 
analysis of forced magnetohydrodynamic waves 
in the inertia-free limit, where the ratio 
of nonlinear inertia to Coriolis
forces is small 
not only
on the core depth but also on the length 
scale of
convection. It is shown that
slow MAC waves are essential for the generation of 
strong polar vortices 
in dipole-dominated dynamos.

In Section \ref{linmodel}, a simplified 
linear model of the
TC is considered wherein a buoyancy disturbance
evolves in an unstably stratified rotating fluid subject to 
an axial magnetic field. It is shown that the intensity
of the slow MAC waves would be at least as high as that
of the fast waves for $Le \sim 0.1$, where $Le$
measures the ratio of Alfv\'en wave to
inertial wave frequencies.
 Section \ref{nonlin} investigates
TC convection in spherical dynamo simulations
at progressively increasing forcing spanning the
dipole-dominated regime up to the onset of polarity
reversals.  The measurement of
wave motions within the TC enables 
a comparison of the dynamics with that predicted
by the simplified linear model.
The main results of this paper are summarized in
Section \ref{concl}.
\section{A tangent cylinder model: 
Evolution of an isolated buoyancy disturbance}
\label{linmodel}
\begin{figure}
	\centering
	\includegraphics[width=0.5\linewidth]{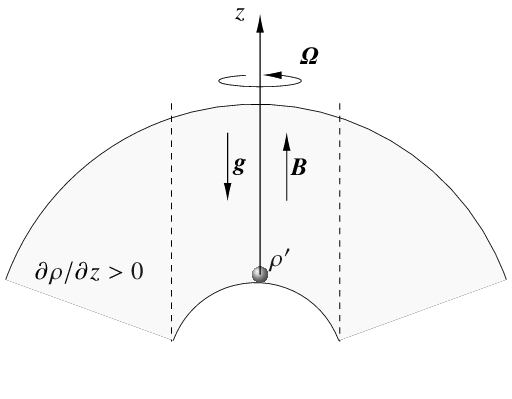}
	\caption{A density perturbation $\rho^\prime$ evolves in an unstably
stratified fluid layer under a uniform axial magnetic field $\bm{B}$
and background rotation $\bm {\varOmega}$.}
	\label{setup}
\end{figure}

Since the dominant magnetic field within the tangent cylinder
is known to be axial, we consider the evolution of a density
perturbation $\rho^\prime$ under 
gravity $\bm{ g}=-g\  \hat{\bm{e}}_z$,
a uniform axial magnetic field $\bm{B}= B \hat{\bm{e}}_z$
and background rotation $\bm{\varOmega }= \varOmega \hat{\bm{e}}_z$
(figure \ref{setup}). In
cylindrical polar coordinates $(s,\phi,z)$, this perturbation is symmetric
about its own axis, or in other words, independent of $\phi$.
Since
$\rho^\prime$ is related to a 
temperature perturbation $\theta$ by $\rho^\prime = - \rho
\alpha \theta$, where $\rho$ is the ambient density and
$\alpha$ is the coefficient of thermal expansion, an initial
temperature perturbation is chosen in the form
\begin{equation}
\theta_0 =  A \exp \left [- 2(s^2+z^2)/\delta^2 \right],
\label{ic}
\end{equation} 
where $A$ is a constant and $\delta$ is the length scale of the perturbation.
\subsection{Governing equations and solutions}
\label{problem_setup}
 In the Boussinesq approximation, the following
linearized MHD equations describe the evolution of
$\bm{u}$, $\bm{b}$ and $\theta$:
\begin{eqnarray}
& &\frac{\partial{\bm{u}}}{\partial{t}}=-\frac{1}{\rho}\nabla{p^*}-2\bm{\varOmega}\times \bm{u}+
\frac{1}{\mu \rho}\left(\bm{B} \cdot \nabla \right) \bm{b} - \bm{g} \alpha\theta +\nu\nabla^2\bm{u},
\label{mom}\\
& &\frac{\partial{\bm{b}}}{\partial{t}} =\left(\bm{B} \cdot \nabla \right) {\bm{u}}
+\eta \nabla^2 \bm{b},
\label{ind}\\
&& \frac{\partial{\theta}}{\partial{t}}=
-\beta \bm{\hat{e}}_z \cdot \bm{u}+ \kappa\nabla^2\theta,
\label{temp}\\
&& \nabla \cdot \bm{u} = \nabla \cdot \bm{b}=0,
\label{divcond}
\end{eqnarray}
where $\nu$ is the kinematic viscosity,
$\kappa$ is the thermal diffusivity, $\eta$ is the magnetic diffusivity,
$\mu$ is the magnetic permeability,  $p^*=p-(\rho/2)|\bm{\varOmega}\times\bm{x}|^2+ (\bm{B}\cdot\bm{b})/\mu$ and 
$\beta=\partial T_0/\partial z<0$ is the mean axial temperature gradient
in the unstably stratified fluid.

In a quiescent fluid,
the initial velocity $\bm{u}_0 $ is zero, 
and since the magnetic field
perturbation takes finite time to develop by induction, 
the initial induced field
$\bm{b}_0$ is also zero. The initial temperature perturbation \eqref{ic}
produces a poloidal flow which interacts with 
$\bm{\varOmega}$ to generate
a toroidal flow, so that the instantaneous state of the flow is defined by
\citep{jfm21}

\begin{equation}\label{utp}
	\bm{u}=u_\phi\hat{\bm{e}}_\phi+\nabla\times[(\psi/s)\hat{\bm{e}}_\phi],
\end{equation}
\begin{equation}\label{lap1}
\nabla_*^2 \psi=\frac{\partial^2 \psi}{\partial z^2}
+s\frac{\partial}{\partial s}\left(\frac{1}{s}\frac{\partial \psi}{\partial s}\right)
=-s \zeta_\phi,
\end{equation}
where $\psi$ is the Stokes streamfunction of the velocity and 
$\bm{\zeta}$ is the vorticity.
Likewise, the instantaneous state of the induced magnetic field is defined by,
\begin{equation}\label{btp}
	\bm{b}=b_\phi\hat{\bm{e}}_\phi+\nabla\times[(\xi/s)\hat{\bm{e}}_\phi],
\end{equation}
\begin{equation}\label{lap2}
\nabla_*^2 \xi=-s\mu j_\phi,
\end{equation}
where $\xi$ is the Stokes streamfunction of the induced magnetic 
field and $\bm{j}$ is the electric current density.

Algebraic simplifications of the governing equations
give an equation for the evolution of $\psi$ in
the form \citep{jfm21}
\begin{equation}
\begin{aligned}
\bigg[ & \left(\frac{\partial{}}{\partial{t}}-
\nu\nabla_*^2\right)\left(\frac{\partial{}}{\partial{t}}
-\eta\nabla_*^2\right)  -V_M^2\frac{\partial^2}{\partial{z^2}}\bigg]^2
\left(\frac{\partial{}}{\partial{t}}-\kappa\nabla_*^2\right)
\left(\nabla_*^2\psi\right)\\
= & - 4 \varOmega^2 \, \left(\frac{\partial{}}{\partial{t}}-
\kappa\nabla_*^2\right)\left(\frac{\partial{}}{\partial{t}}-
\eta\nabla_*^2\right)^2 \frac{\partial^2{\psi}}{\partial{z^2}}\\
& - g \alpha \beta s \, \left(\frac{\partial{}}{\partial{t}}-\eta\nabla_*^2\right)
\left[\left(\frac{\partial{}}{\partial{t}}-
\nu\nabla_*^2\right)\left(\frac{\partial{}}{\partial{t}}-
\eta\nabla_*^2\right)-V_M^2
\frac{\partial^2}{\partial{z^2}}\right]
\frac{\partial{}}{\partial{s}}\left(\frac{1}{s}
\frac{\partial{\psi}}{\partial{s}}\right), \label{psieq}
\end{aligned}
\end{equation}
where $V_M= B/\sqrt{\mu \rho}$ is the Alfv\'en wave velocity.

By applying the Hankel--Fourier transform 
\begin{equation}
H_1 \{\psi(s,z)/s\} = \hat{\psi}(k_s,k_z) = 
\frac{1}{2\pi^2}\int_{0}^{\infty}
\int_{0}^{\infty}\psi(s,z) J_1(k_s s) \mbox{e}^{-ik_z z}
\,\mbox{d}s\,\mbox{d}z
\label{hk1}
\end{equation}
to \eqref{psieq}, where $J_1$ is the first-order Bessel
function of the first kind, we obtain,
\begin{equation}
\begin{aligned}
\bigg[ & \left(\frac{\partial{}}{\partial{t}}+ \nu k^2 \right)
\left(\frac{\partial{}}{\partial{t}}+ \eta k^2 \right) + 
V_M^2k_z^2\bigg]^2\left(\frac{\partial{}}{\partial{t}} + \kappa k^2 \right)\hat{\psi}\\
= & - \frac{4 \varOmega^2k_z^2}{k^2}\left(\frac{\partial{}}{\partial{t}}+
\eta k^2\right)^2\left(\frac{\partial{}}{\partial{t}}+\kappa k^2\right)\hat{\psi}\\
& -  \frac{g \alpha \beta k_s^2}{k^2}\left(\frac{\partial{}}{\partial{t}}+
\eta k^2\right)\left[\left(\frac{\partial{}}{\partial{t}}+
\nu k^2\right)\left(\frac{\partial{}}{\partial{t}}+
\eta k^2\right)+V_M^2k_z^2\right]\hat{\psi}.
\end{aligned}
\label{psiwave}
\end{equation}
Considering plane wave solutions of the 
form $\hat{\psi} \sim  \mbox{e}^{i\lambda t}$,
we obtain the relation
\begin{equation}
\begin{aligned}
\lambda^5&-2i\omega_\eta\lambda^4-
\left(\omega_A^2+\omega_\eta^2+2\omega_M^2+
\omega_C^2\right)\lambda^3+2i\omega_\eta\left(\omega_A^2+
\omega_M^2+\omega_C^2\right)\lambda^2\\&+
\left(\omega_A^2\omega_\eta^2+\omega_A^2\omega_M^2+
\omega_M^4+\omega_\eta^2\omega_C^2\right)\lambda-
i\omega_A^2\omega_\eta\omega_M^2=0.
\label{chareqn2}
\end{aligned}
\end{equation}
for a system where both viscous
and thermal diffusion are much smaller 
than magnetic diffusion ($\nu,\kappa \to 0$).

The fundamental frequencies in \eqref{chareqn2} are given by,
\begin{subequations}\label{eq:define_omega}    
	\begin{gather}  
		\omega_C^2=\frac{4\varOmega^2 k_z^2}{k^2},		\quad
		 	\omega_M^2=V_M^2 k_z^2,	
\quad	 \omega_A^2=g\alpha\beta \frac{k_s^2}{k^2},	
\quad\omega_\eta^2=\eta^2 k^4,\tag{\theequation a-d} 
	\end{gather}
\end{subequations}

representing linear inertial waves, Alfv\'en waves,
internal gravity waves and magnetic diffusion respectively. 
In unstable density stratification, $\omega_A^2<0$. Here, $k^2=k_s^2+k_z^2$.

 For the frequency inequality
 $|\omega_C|\gg|\omega_M|\gg|\omega_A|\gg|\omega_\eta|$, 
the roots of equation \eqref{chareqn2} are approximated
by \citep{jfm21}
\begin{eqnarray}
\lambda_{1,2} &\approx& \pm \left(\omega_C+ \frac{\omega_M^2}{\omega_C} \right) +
i \, \frac{\omega_M^2\omega_\eta}{\omega_C^2}, \label{l12approx}\\ 
\lambda_{3,4} &\approx& \pm \left(\frac{\omega_M^2}{\omega_C}+
\frac{\omega_A^2}{2\omega_C}\right) + i\, \omega_\eta \,
\left(1-\frac{\omega_A^2}{2\omega_M^2}\right), \label{l34approx} \\
\lambda_{5} &\approx& i \, \frac{\omega_A^2\omega_\eta}{\omega_M^2}, \label{l5approx}
\end{eqnarray}
representing damped fast MAC waves ($\lambda_{1,2}$), damped
slow MAC waves ($\lambda_{3,4}$), and the overall growth
of the perturbation ($\lambda_5$).

The general solutions for $\hat{\psi}$, $\hat{u_\phi}$, 
$\hat{\xi}$, $\hat{b_\phi}$ are then given by 
\begin{equation}\label{gensol}
	\begin{aligned}
		[\hat{\psi},\hat{u_\phi},\hat{\xi},\hat{b_\phi}]
		=\sum_{m=1}^{5}[D_m,G_m,P_m,Q_m]\mbox{e}^{i\lambda_m t},
	\end{aligned}
\end{equation}
where the coefficients $D_m$, $G_m$, $P_m$ and $Q_m$ 
are evaluated from the 
initial conditions for $\hat{\psi}$, 
$\hat{u_\phi}$, $\hat{\xi}$, $\hat{b_\phi}$ 
and their time derivatives. 
The poloidal velocity is obtained from
its streamfunction using the relations
\begin{subequations}\label{polvel}    
	\begin{gather}  
		\hat{u}_z=k_s\hat{\psi},\quad \hat{u}_s
		=-i k_z\hat{\psi}.\tag{\theequation a-b} 
	\end{gather}
\end{subequations}

\subsection{MAC waves in unstable stratification}
\label{linresults}
The analysis of the solutions is limited to times much shorter than the
time scale for the exponential increase of the perturbations. 
When the buoyancy force is small
compared with the Lorentz force ($|\omega_A/\omega_M| \ll 1$), 
the parameter regime is determined by the
Lehnert number $Le$ and the magnetic Ekman number $E_\eta$,
\begin{subequations}\label{ledef}    
	\begin{gather}  
		Le=\frac{V_M}{2\varOmega\delta},\quad E_\eta
		=\frac{\eta}{2 \varOmega \delta^2},\tag{\theequation a-b} 
	\end{gather}
\end{subequations}
based on the length scale of the initial
perturbation \eqref{ic}.

\begin{figure}
	\centering
	\hspace{-1.2 in}	(a)  \hspace{1.4 in} (b)  \hspace{1.4 in} (c) \\
	\includegraphics[width=0.25\linewidth]{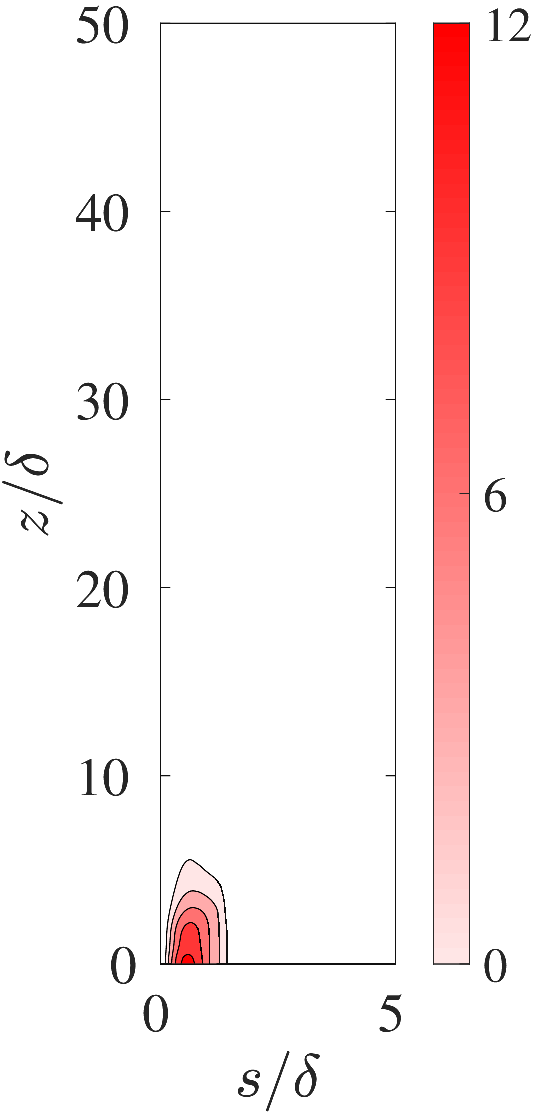}
	\includegraphics[width=0.25\linewidth]{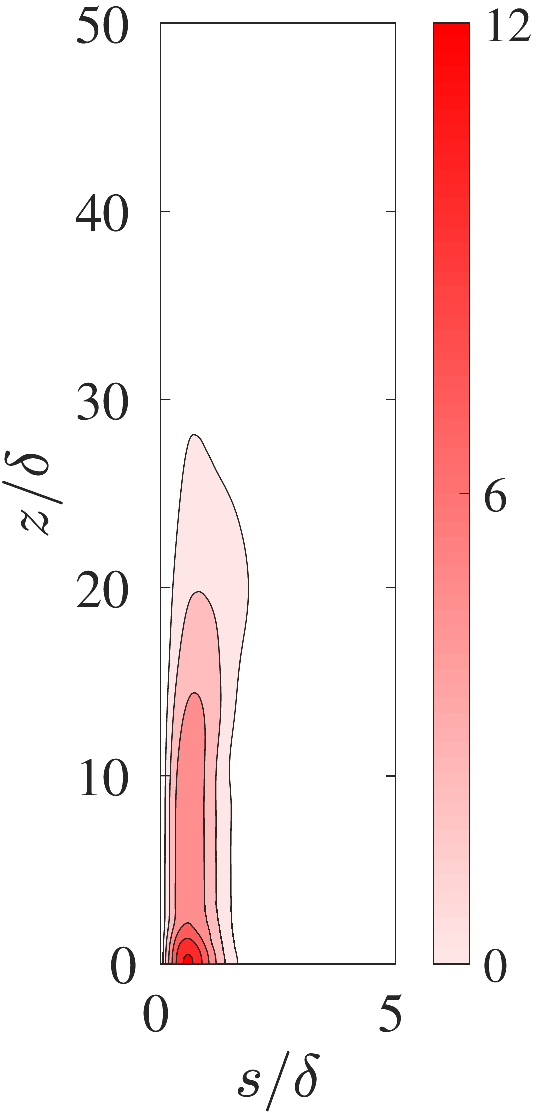}
	\includegraphics[width=0.25\linewidth]{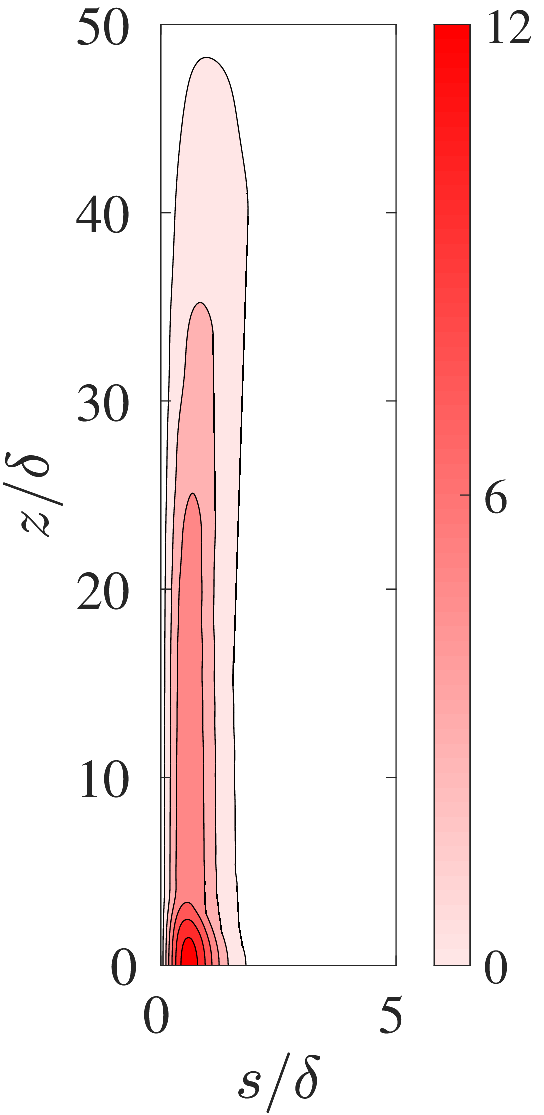}	
	\caption{Evolution of the poloidal velocity streamfunction 
$\psi$ with time 
		(measured in units of the magnetic diffusion time $t_\eta$) 
		for $Le=0.09$ and $E_\eta = 2\times10^{-5}$. The snapshots 
		are at (a) $t/t_\eta = 1\times10^{-3}$, (b) $t/t_\eta = 5\times10^{-3}$ 
		and (c) $t/t_\eta = 1\times10^{-2}$. 
The ratio $|\omega_A/\omega_M|= 0.05$ 
		at times after the formation of the waves.}
	\label{psi1}
\end{figure}

Figure \ref{psi1} shows the evolution in time of the 
poloidal velocity streamfunction, obtained from the  
inverse  Hankel--Fourier transform
\begin{equation}\label{hftrans}
	\begin{aligned}
		\psi(s,z)=
		4 \pi s\int_{0}^{\infty}\int_{0}^{\infty}
\hat{\psi}(k_s,k_z)J_1(k_s s)\mbox{e}^{i k_z z} k_s dk_sdk_z,
	\end{aligned}
\end{equation}
computed by setting the upper limits of the integrals 
(the truncation values of $k_s$
and $k_z$) to $3/\delta$. The formation of a columnar 
 structure from the initial perturbation through wave motions is evident.

\begin{figure}
	\centering
	\hspace{-2.5 in}	(a)  \hspace{2.5 in} (b) \\
	\includegraphics[width=0.4\linewidth]{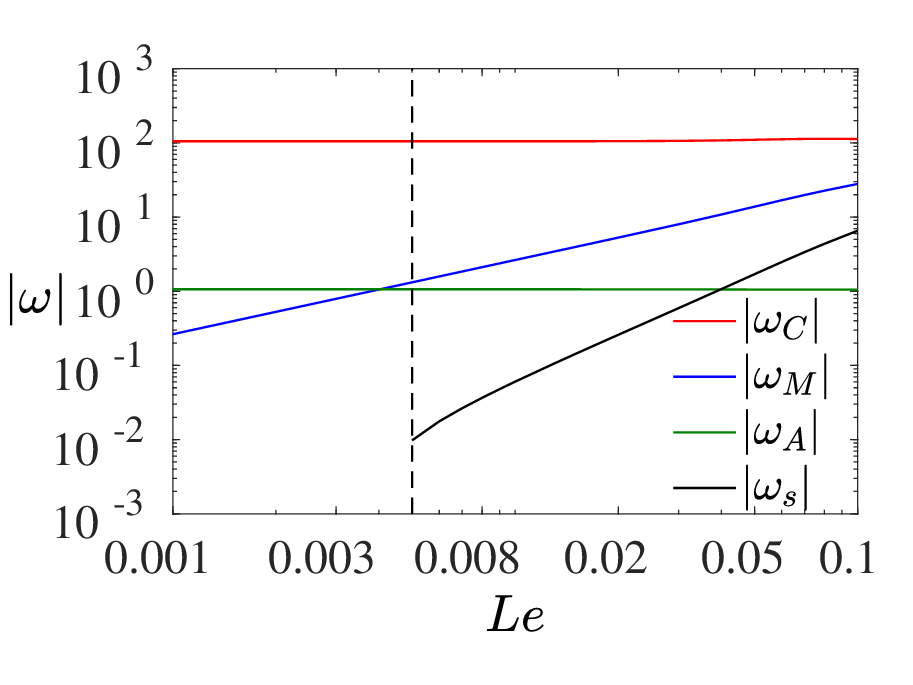}
	\includegraphics[width=0.4\linewidth]{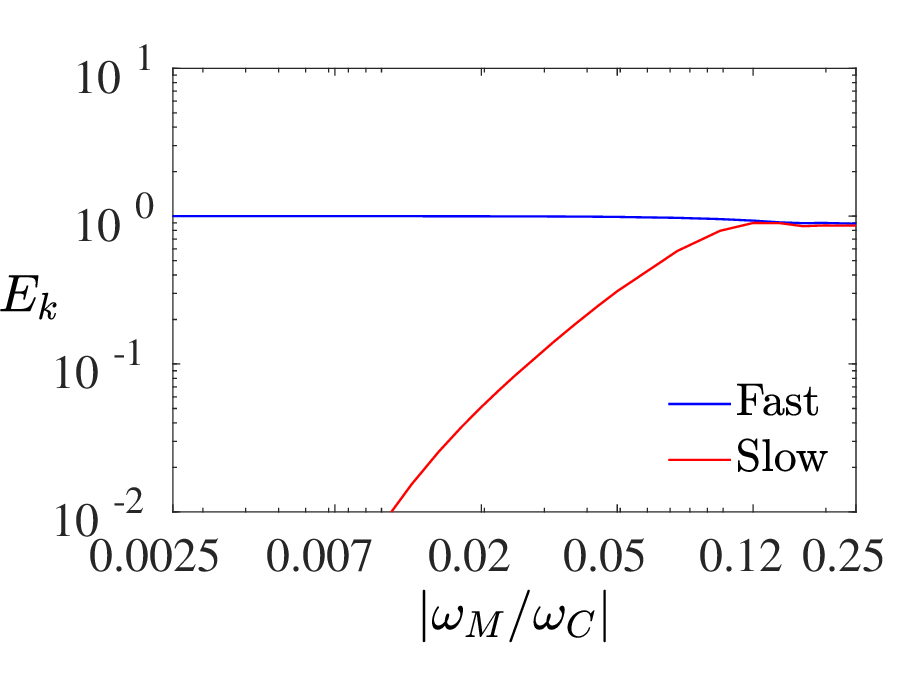}\\
	\hspace{-2.5 in}	(c)  \hspace{2.5 in} (d) \\
	\includegraphics[width=0.4\linewidth]{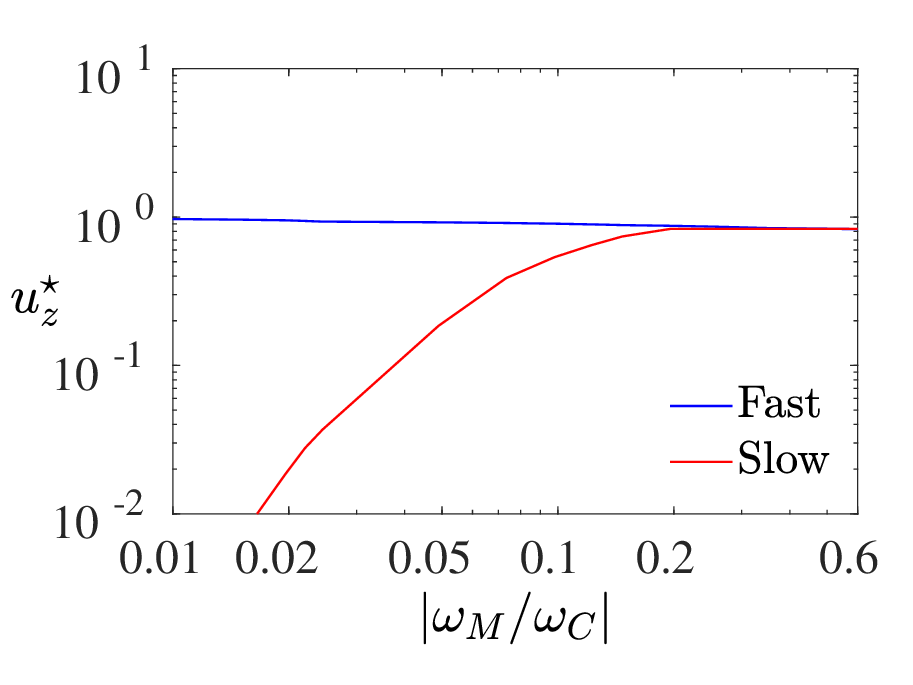}
	\includegraphics[width=0.4\linewidth]{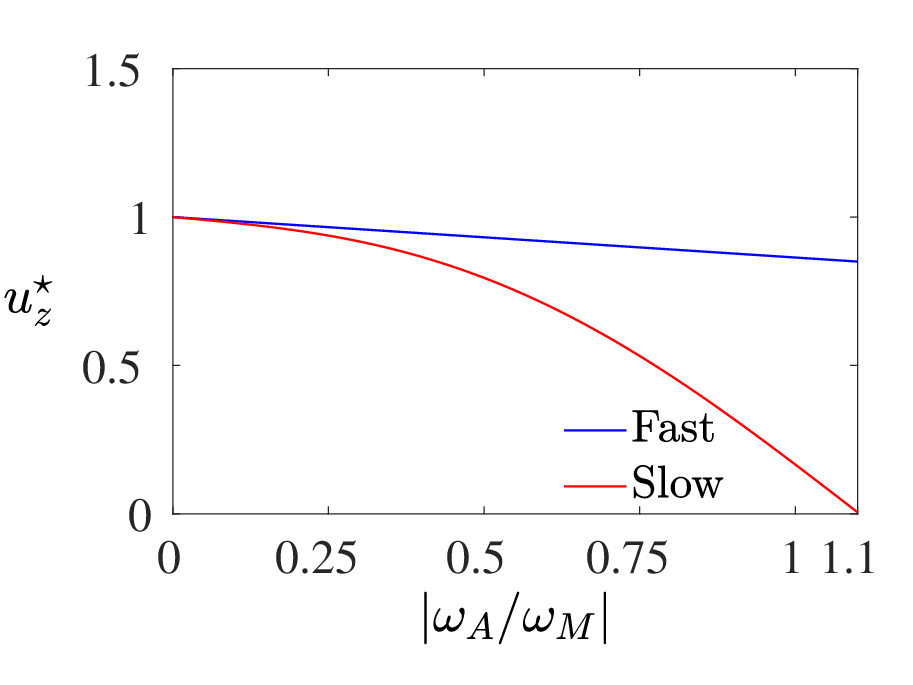}\\
	\caption{(a) Variation of absolute values of the
frequencies with 
Lehnert number $Le$. 
		(b) Variation of the  kinetic energy of fast and slow MAC 
waves (normalized by the nonmagnetic kinetic energy)
 with $|\omega_M/\omega_C|$. 
The range of $|\omega_M/\omega_C|$ in 
		(b) corresponds to the range of $Le$ in (a).
Here, $|\omega_A/\omega_M|=0.1$. 
(c) Variation of the peak values of the fast and slow wave 
parts of the $z$ velocity (normalized by its
nonmagnetic value) with $|\omega_M/\omega_C|$.
(d) Variation of the peak $z$ velocity of fast and slow MAC
waves with $|\omega_A/\omega_M|$ for $Le=0.09$. 
All calculations are 
		performed for 
$E_\eta=2\times10^{-5}$ and $t/t_\eta=5\times10^{-3}$.}
	\label{linplots}
\end{figure}

In figure \ref{linplots} (a), the fundamental frequencies are
based on the mean wavenumbers obtained from
ratios of 
$L^2$ norms,
\begin{subequations}\label{l2norm}    
	\begin{gather}  
		\bar{k}_s=\frac{||k_s \hat{\psi} k||}{||\hat{\psi}k||},
\quad \bar{k}_z
		=\frac{||k_z \hat{\psi} k||}{||\hat{\psi} k||},
\quad \bar{k}=
		\frac{||\hat{\psi}k^2||}{||\hat{\psi}k||}.
\tag{\theequation a-c} 
	\end{gather}
\end{subequations}

For $Le>0.005$, the inequality 
$|\omega_C| > |\omega_M| > |\omega_A| $
exists, indicating the MAC wave regime.
To obtain the relative intensities of the fast and slow MAC
waves, the fast and slow MAC wave parts of the general solution
are separated, as in earlier studies \citep{jfm17b,jfm21}.

The kinetic energy is calculated using the Parseval's theorem,
	\begin{equation}
		E_k=16\pi^4\int_{0}^{\infty}
		\int_{0}^{\infty}\left(\hat{u}_{s}^2+\hat{u}_{z}^2
		+\hat{u}_{\phi}^2\right) k_s \mbox{d} k_s \mbox{d} k_z,  
		 \label{eq:ekf}
	\end{equation}
separately for the fast and slow MAC waves, with
the upper limits of the integrals 
set to $10/\delta$ in the computations.

\begin{figure}
	\centering
	\hspace{-1.2 in}	(a)  \hspace{1.4 in} (b)  \hspace{1.4 in} (c) \\\vspace{0.1in}
	
	\includegraphics[width=0.25\linewidth]{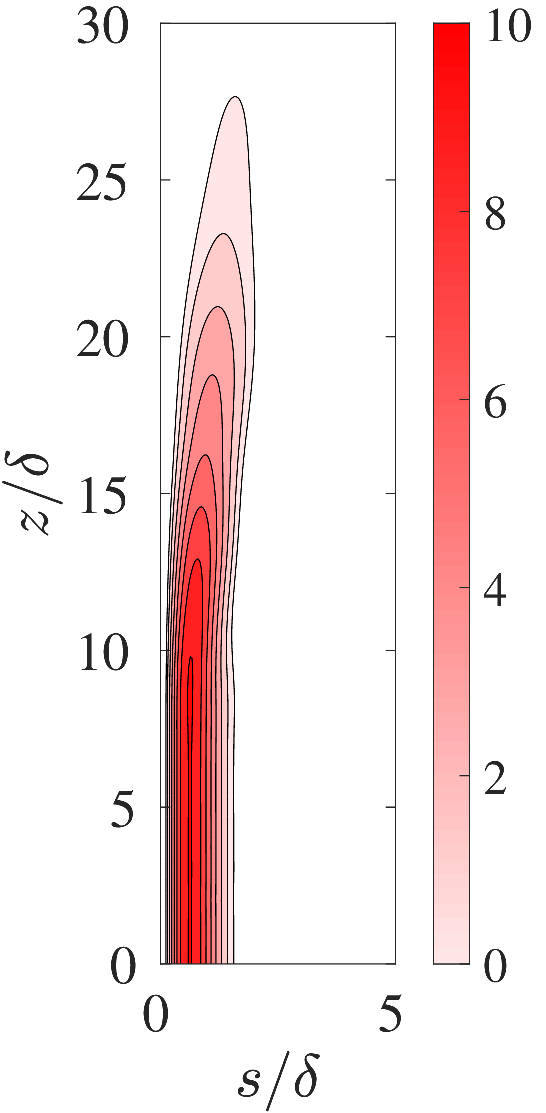}
	\includegraphics[width=0.25\linewidth]{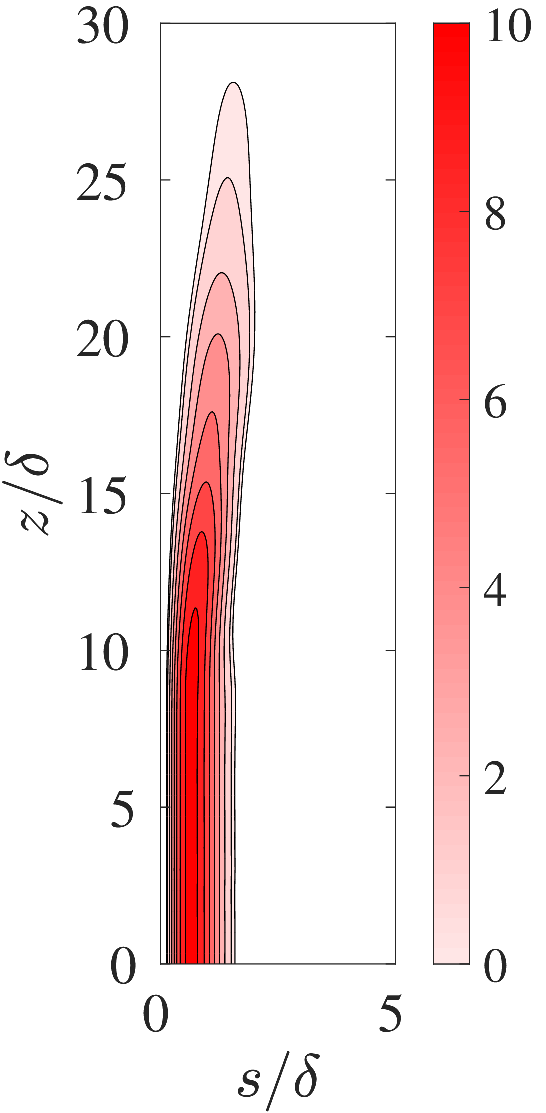}
	\includegraphics[width=0.25\linewidth]{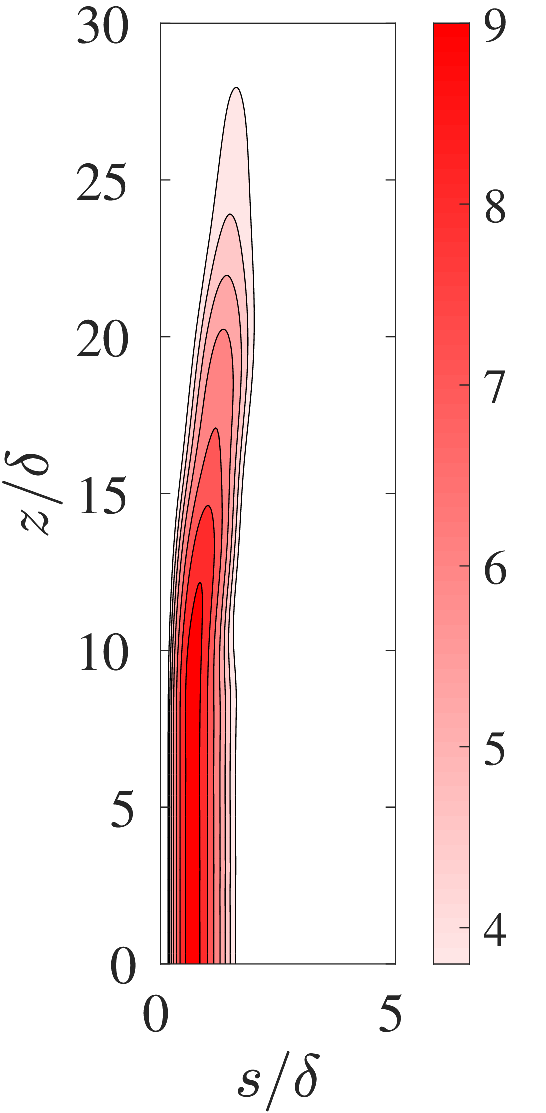}\\
	\hspace{-1.2 in}	(d)  \hspace{1.4 in} (e)  
\hspace{1.4 in} (f) \\\vspace{0.1in}
	\includegraphics[width=0.25\linewidth]{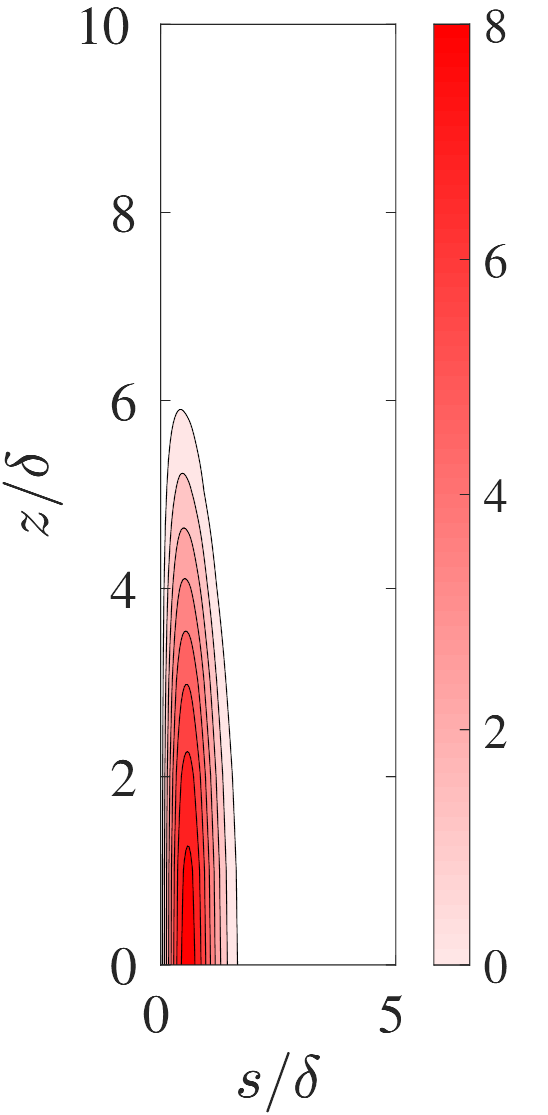}
	\includegraphics[width=0.25\linewidth]{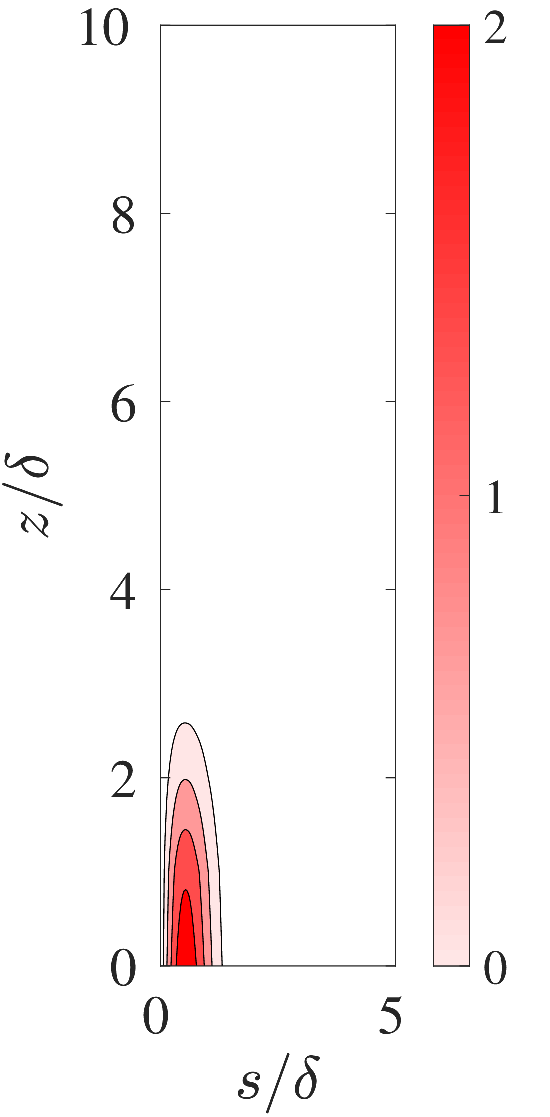}
	\includegraphics[width=0.25\linewidth]{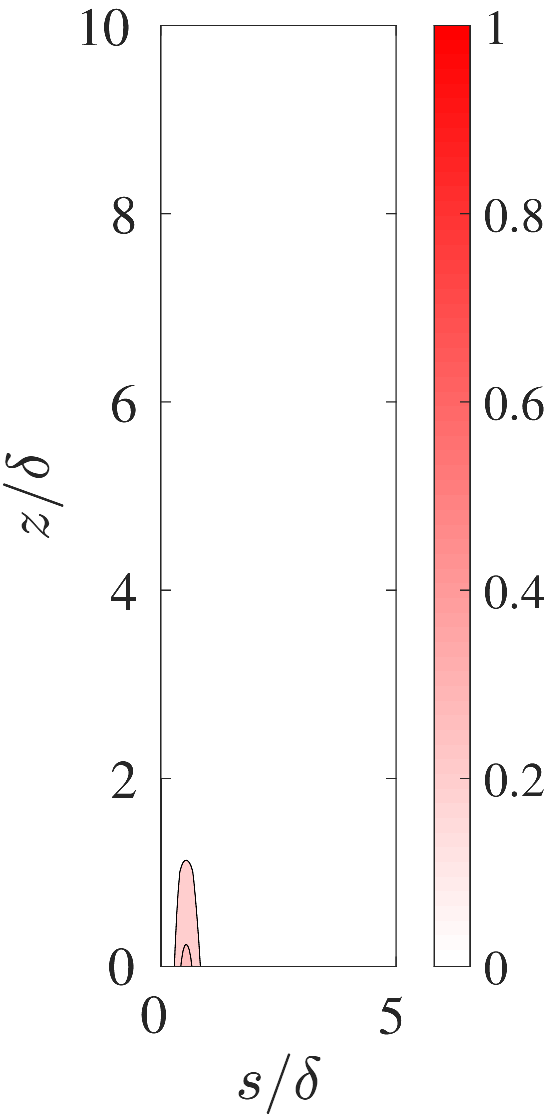}\\
	\caption{Streamfunction $\psi$ of the fast wave part of the
solution ((a)--(c)), and that for the slow wave part ((d)--(f)) for 
		$|\omega_A/\omega_M|=0.1,0.6,0.95$ (shown left to right). 
		The plots are generated at time $t/t_\eta = 5\times10^{-3}$ 
		for the parameters $Le=0.09$ and $E_\eta = 2\times10^{-5}$.}
	\label{psi2}
\end{figure}

Figure \ref{linplots} (b) shows the variation of the
total kinetic energy $E_k$
of the two waves with $|\omega_M/\omega_C|$. 
The range of $|\omega_M/\omega_C|$ in 
		figure \ref{linplots}
(b) corresponds to the range of $Le$ in figure \ref{linplots} (a).
Both 
$Le$ and  $|\omega_M/\omega_C|$ are of the same order of
magnitude when the 
 magnetic field and rotation axes are parallel \citep{jfm21}. 
(Since the initial wavenumber $k_0=\sqrt{6}/\delta$,
$|\omega_M/\omega_C|= \sqrt{6} \, Le$.)
The slow MAC waves appear when $\omega_M$ exceeds $\omega_A$,
  and their energy becomes comparable to 
that of the fast waves for $Le\sim 0.1$. Figure \ref{linplots} (c)
shows that the peak value of the slow wave
$z$ velocity is approximately equal 
 to that of fast waves
when $|\omega_M/\omega_C| \approx 0.2$, where 
$u_z = s^{-1} \, \partial \psi/\partial s$.
Figure \ref{linplots} (d) shows the 
peak value of $u_z$ for 
both fast and slow waves against $|\omega_A/\omega_M|$ 
at $Le=0.09$ ($|\omega_M/\omega_C|=0.22$). The slow wave
velocity decreases appreciably and tends to zero as 
$|\omega_A/\omega_M|\approx 1$. The fast wave velocity, on the
other hand, does not change much with increasing buoyancy.
Figures \ref{linplots} (a)--(d) indicate that
the intensity of slow MAC wave motions would be comparable to
that of the fast waves in the regime thought to be relevant to
dipole-dominated dynamos, $|\omega_M/\omega_C| \sim 0.1$ and
$|\omega_A/\omega_M| < 1$. In the 
regime $|\omega_A/\omega_M| \sim 1$, where dipole collapse
is found to happen \citep{jfm23}, the slow MAC wave velocity
goes to zero. 
The effect of increasing buoyancy on the fast and slow MAC waves
are shown graphically in figure \ref{psi2}.
Here, the magnitudes of the fast and slow waves are comparable at 
$|\omega_A/\omega_M|= 0.1$, whereas the slow waves are severely
attenuated for $|\omega_A|\sim|\omega_M|$. The fast waves
are practically unaffected by the increased forcing.

\begin{figure}
	\centering
	\includegraphics[width=0.45\linewidth]{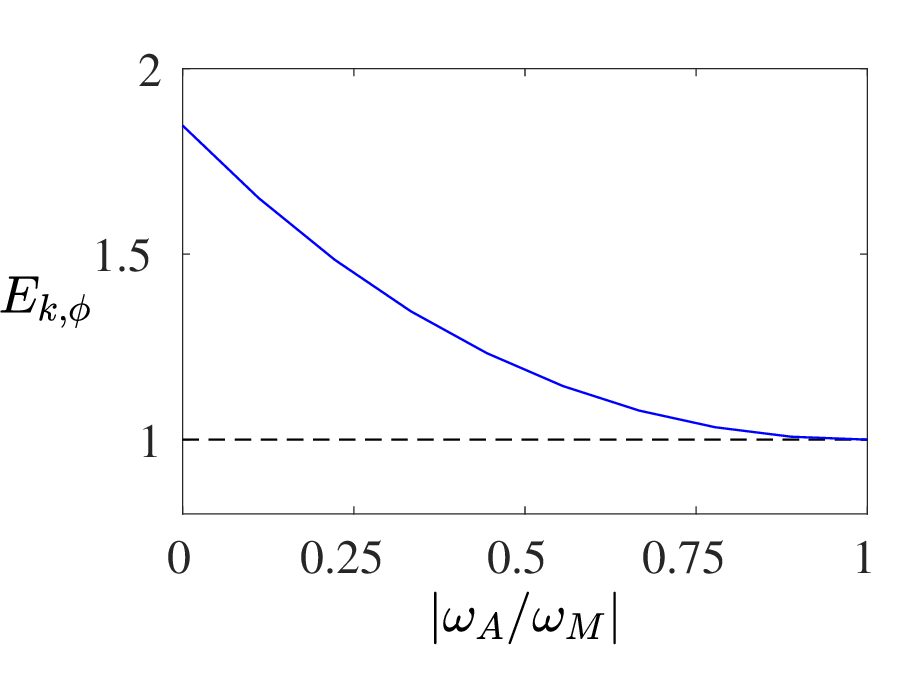}
	\caption{ Variation with $|\omega_A/\omega_M|$
of the $\phi$
component of the kinetic energy, $E_{k,\phi}$ normalized by its
 nonmagnetic value.
		The calculations are 
		performed for 
		$E_\eta=2\times10^{-5}$, $t/t_\eta=5\times10^{-3}$ and
$|\omega_M/\omega_C|=0.225$.}
	\label{uphi}
\end{figure}

Figure \ref{uphi} shows the variation of the $\phi$ component 
of kinetic energy, normalized by its nonmagnetic
value, with progressively increasing forcing for
a given $\omega_M/\omega_C$. As $|\omega_A/\omega_M|$
tends to unity, the energy
tends to the nonmagnetic
value, consistent with the suppression of the slow waves.
This result has implications for the intensity of
polar vortices in strongly forced dynamos where
$|\omega_A/\omega_M| \sim 1$ within the tangent cylinder.

In Section \ref{nonlin},
we examine whether slow MAC waves are influential in 
the generation of strong polar vortices in dipole-dominated
spherical dynamos. We also
study the regime of dipole collapse, wherein we expect TC
convection to be supported by only the fast waves,
resulting in much weaker polar circulation. 

\section{Nonlinear dynamo simulations}
\label{nonlin}
	We consider an electrically conducting fluid located between two concentric, 
	co-rotating spherical surfaces with a radius ratio of 0.35. These surfaces 
	correspond to the inner core boundary (ICB) and the core-mantle boundary 
	(CMB). Our model is based on the codensity formulation, where thermal 
	and compositional buoyancy are combined \citep{95brag}. 
	The other two body forces are the Lorentz force, which arises from the 
	interaction between the induced electric current and the magnetic field,
	and the Coriolis force, which originates from the background rotation.
	
	In the dynamo models, lengths are scaled by the thickness of 
	the spherical shell $L$ and the 
	time is scaled by magnetic diffusion time $L^2/\eta$, where $\eta$ is 
	magnetic diffusivity. The velocity field $\bm{u}$ is scaled by $\eta/L$, 
	while the magnetic field $\bm{B}$ is scaled by $(2\varOmega\rho\mu\eta)^{1/2}$, 
	where $\varOmega$ represents the rotation rate, $\rho$ represents fluid density, 
	and $\mu$ represents magnetic permeability. The temperature is 
	scaled by $\beta L$, where $\beta$ denotes the radial temperature gradient 
	at the outer boundary.
	
	The non-dimensional equations for velocity, magnetic field, and temperature 
	in MHD under the Boussinesq approximation are given by,

	\begin{align}
E Pm^{-1}  \Bigl(\frac{\partial {\bm u}}{\partial t} + 
(\nabla \times {\bm u}) \times {\bm u}
\Bigr)+  {\hat{\bm{z}}} \times {\bm u} = - \nabla p^\star +
Ra \, Pm Pr^{-1} \, T \, {\bm r} \,  \nonumber\\ +  (\nabla \times {\bm B})
\times {\bm B} + E\nabla^2 {\bm u}, \label{momentum} \\
\frac{\partial {\bm B}}{\partial t} = \nabla \times ({\bm u} \times {\bm B}) 
+ \nabla^2 {\bm B},  \label{induction}\\
\frac{\partial T}{\partial t} +({\bm u} \cdot \nabla) T =  Pm Pr^{-1} \,
\nabla^2 T,  \label{heat1}\\
\nabla \cdot {\bm u}  =  \nabla \cdot {\bm B} = 0.  \label{div}
\end{align}
	
The modified pressure $p^*$  in equation \eqref{momentum}
 is given by $p+\frac{1}{2}E \, Pm^{-1} \, |\bm{u}|^2$.
The dimensionless parameters in the above equations 
are the Ekman number $E=\nu/2\varOmega L^2$,
the Prandtl number, $Pr=\nu/\kappa$; the magnetic Prandtl number, $Pm=\nu/\eta$; 
and the modified Rayleigh number, given by $g\alpha\beta L^2/2\varOmega\kappa$. 
The parameters $g$, $\nu$, $\kappa$, and $\alpha$ denote the 
gravitational acceleration, kinematic viscosity, thermal diffusivity, 
and coefficient of thermal expansion respectively.
	
The temperature distribution in the basic state is described by a basal 
heating profile given by $T_0(r) = r_i r_o/r$, where $r_i$ and $r_o$ 
are the inner and outer radii of the spherical shell respectively. 
The velocity and magnetic fields satisfy the no-slip and electrically 
insulating conditions at their respective boundaries. The inner boundary
 is isothermal, whereas the outer boundary maintains a constant heat flux. 
 The calculations are performed using a pseudospectral code that applies 
 spherical harmonic expansions to the angular coordinates $(\theta, \phi)$ 
 and finite differences to the radius $r$ \citep{willis2007}.

For two values of the Ekman number $E$,
a series of simulations at progressively increasing Rayleigh 
number $Ra$ are performed, spanning
the dipole-dominated regime up to the start of polarity reversals. 
For each $E$, the value of $Pm = Pr$ is chosen such
that the local Rossby number $Ro_\ell$, 
which gives the ratio of the inertial to Coriolis forces on
the characteristic length scale of convection 
\citep{christensen2006scaling} is $< 0.1$ \citep{jfm23}.
Thus, the dynamo simulations lie in the rotationally dominant, or low-inertia, 
regime. For $Pm=Pr$,
the dynamo obtained by solving equations \eqref{momentum}--\eqref{div}
is compared with its nonmagnetic counterpart, obtained by solving
equations \eqref{mom1}--\eqref{div1}, \ref{nmeqns}.

\subsection{TC convection and polar vortices}

\begin{figure}
	\centering
	\hspace{-2.3 in}	(a)  \hspace{1.8 in} (b) \hspace{1.9 in} (c) \\
	\includegraphics[width=0.32\linewidth]{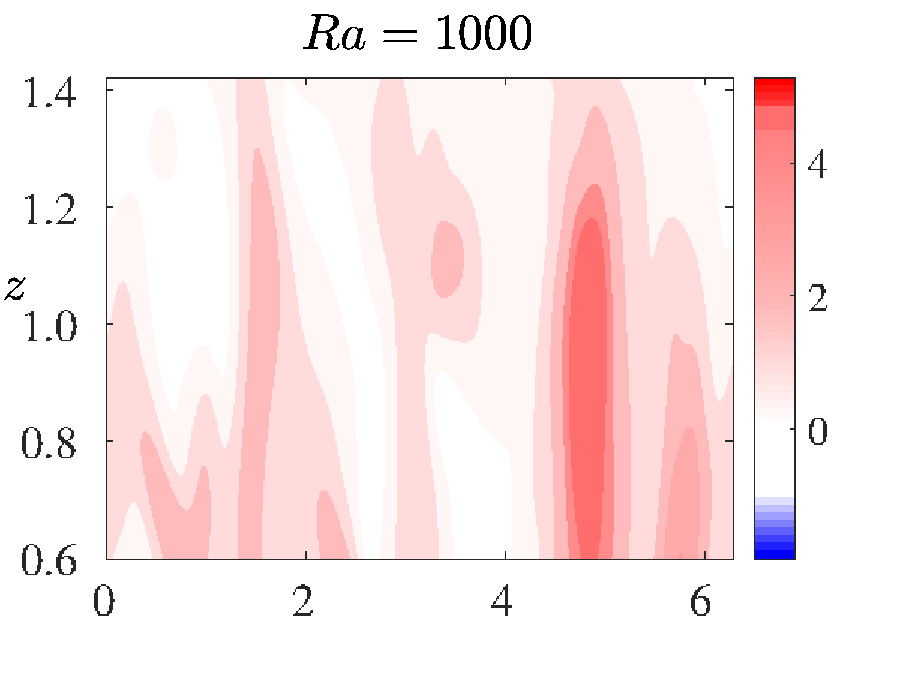}\hspace{0.075in}
	\includegraphics[width=0.32\linewidth]{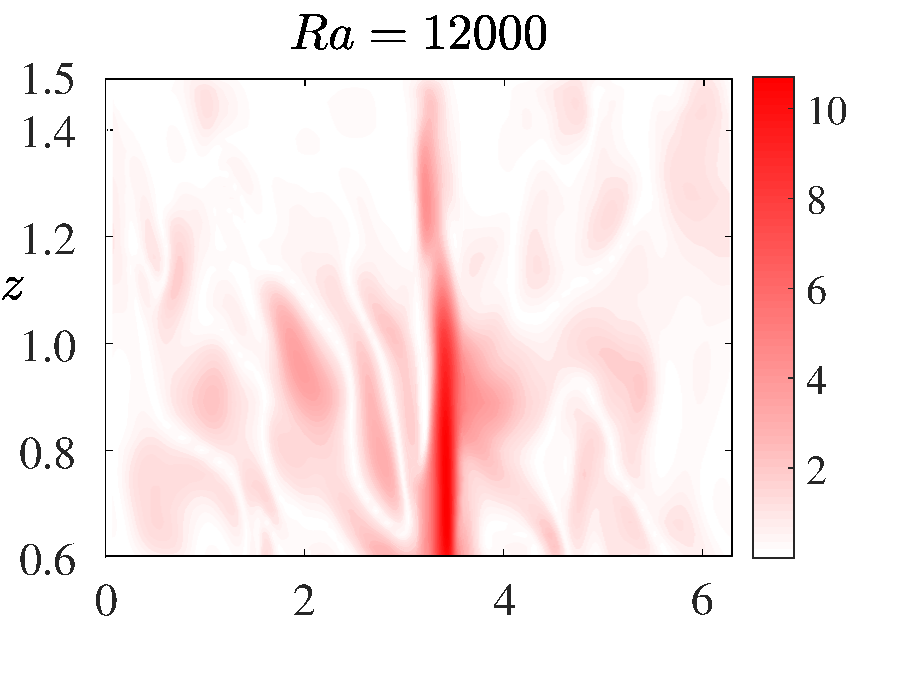}\hspace{0.07in}
	\includegraphics[width=0.32\linewidth]{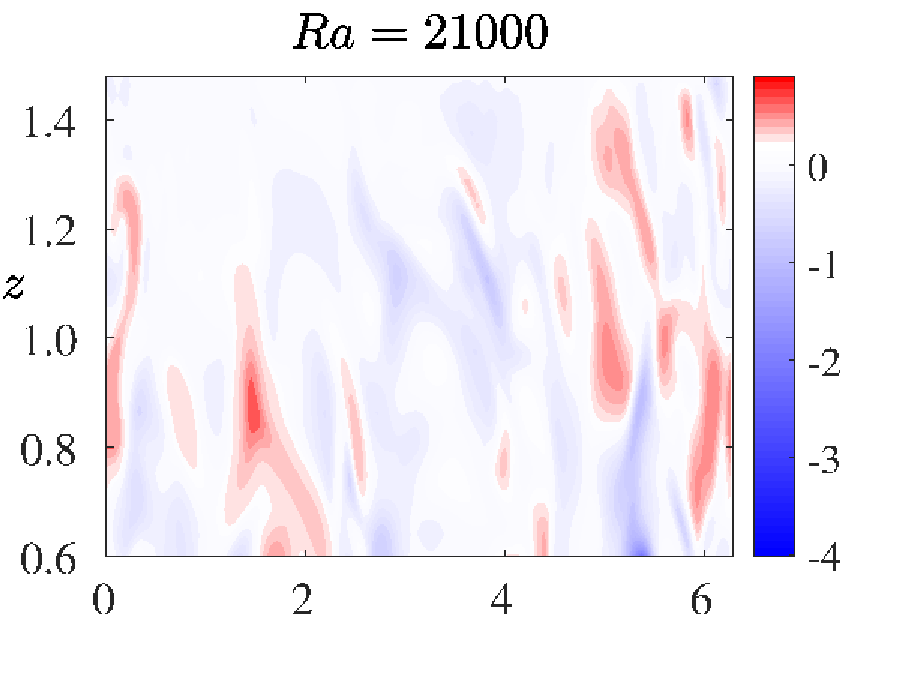}\vspace{-0.2 in}\\
	\hspace{-2.3 in}	(d)  \hspace{1.8 in} (e) \hspace{1.9 in} (f) \\
	\includegraphics[width=0.33\linewidth]{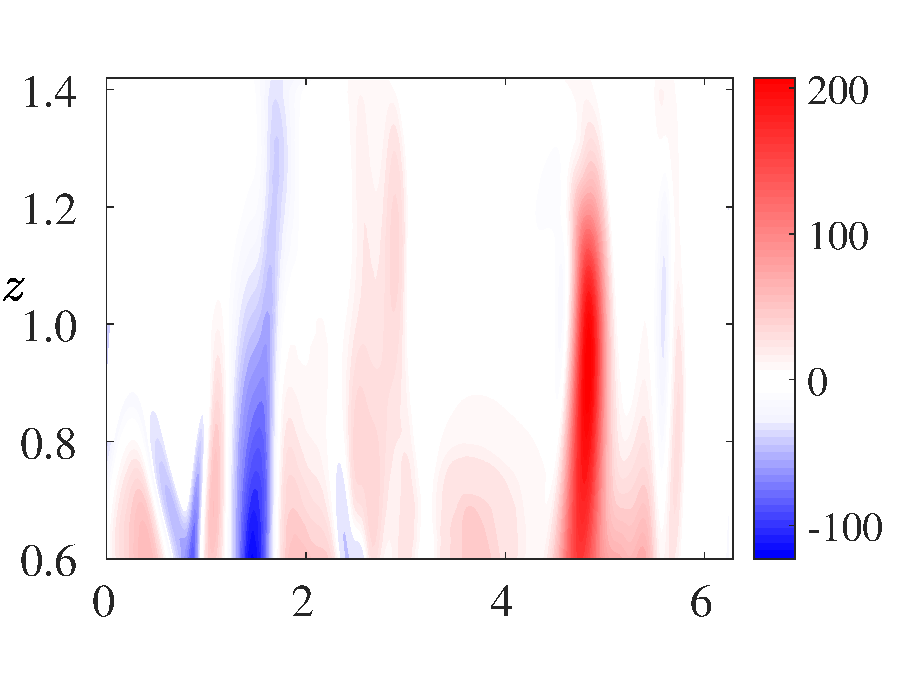}
	\includegraphics[width=0.33\linewidth]{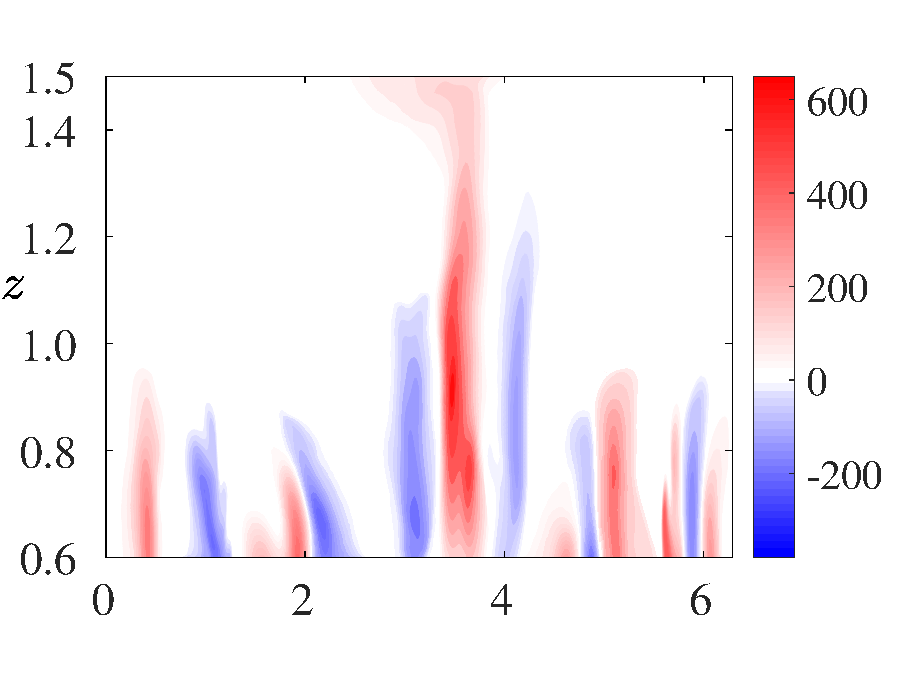}
	\includegraphics[width=0.33\linewidth]{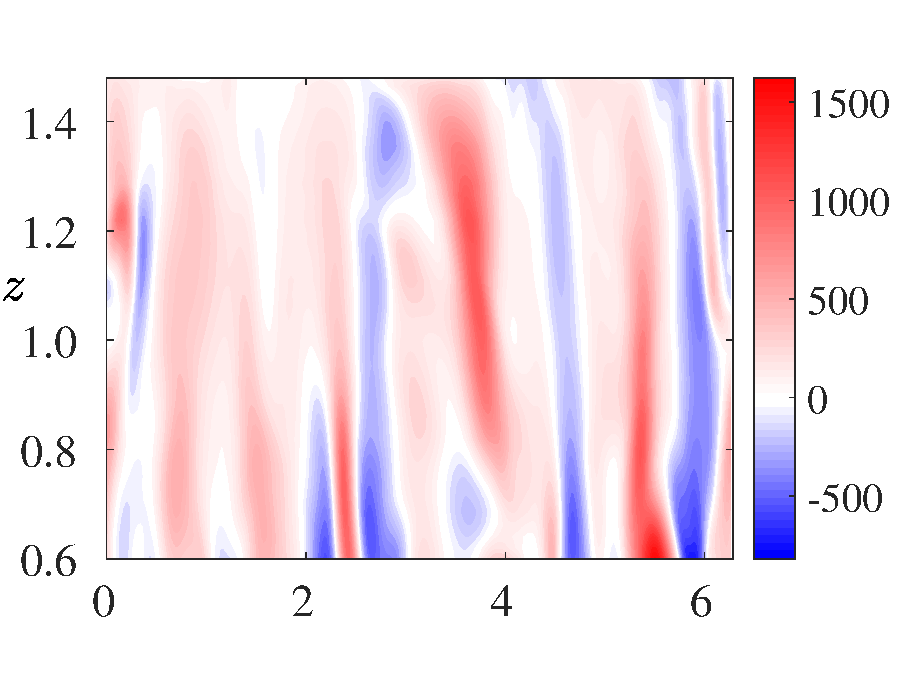}\vspace{-0.2 in}\\
	\hspace{-2.3 in}	(g)  \hspace{1.8 in} (h) \hspace{1.9 in} (i) \\
	\includegraphics[width=0.33\linewidth]{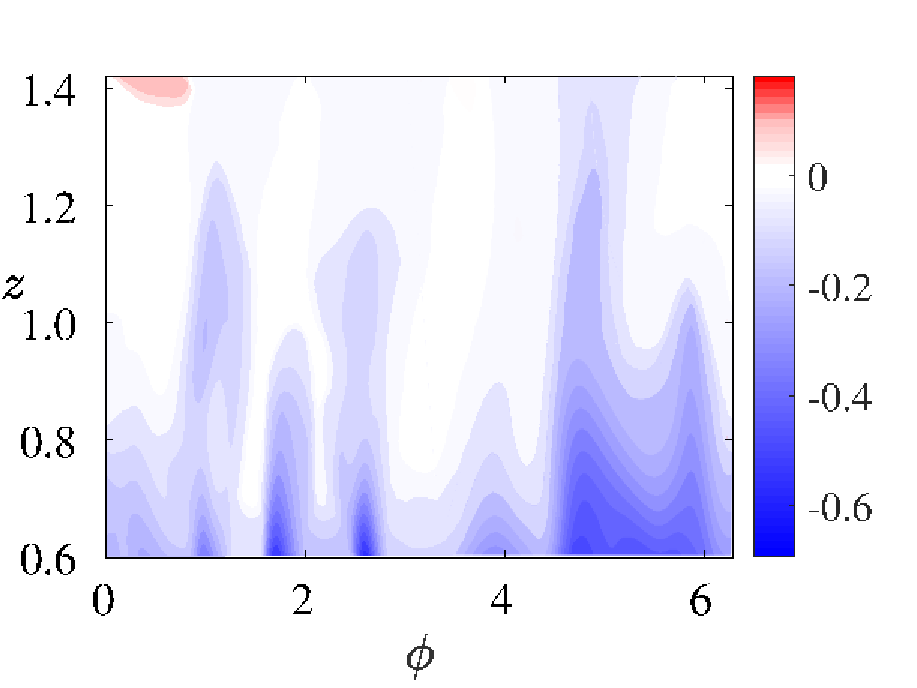}
	\includegraphics[width=0.33\linewidth]{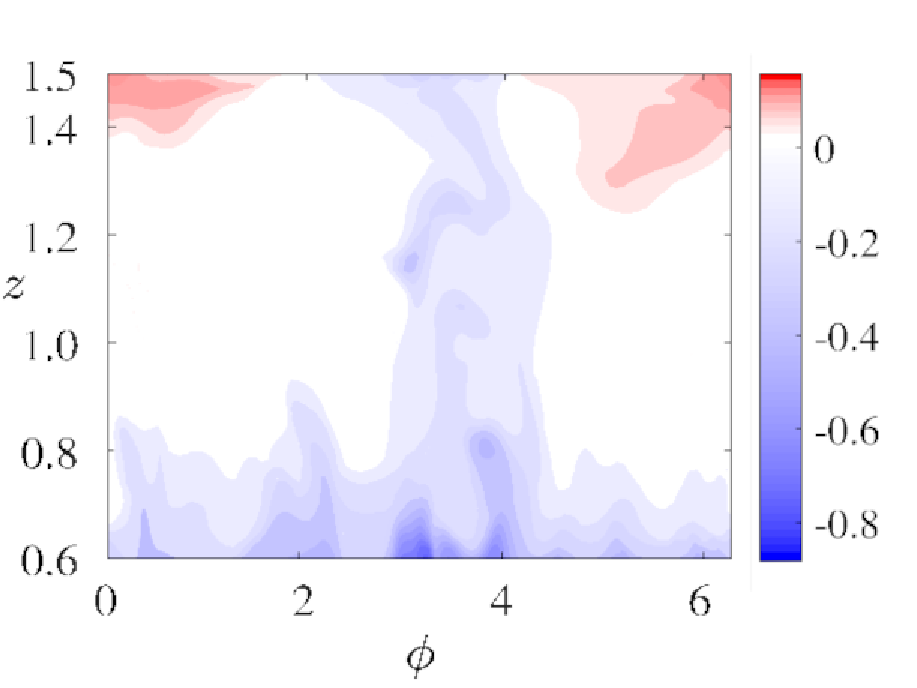}
	\includegraphics[width=0.33\linewidth]{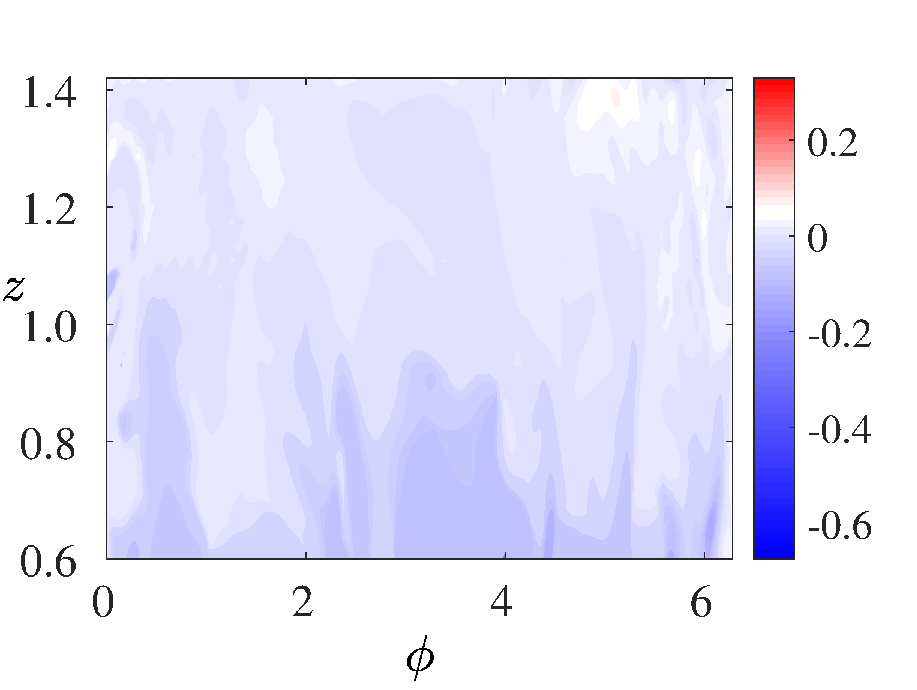}\\
	\caption{Cylindrical section ($z-\phi$) plots within the tangent 
		cylinder of $B_z$ (a-c), $u_z$ (d-f) and $\partial T/\partial z$ (g-i)
		for $E = 6\times 10^{-5}, Pr = Pm = 5$. The figures 
		are shown from the saturated state of the dynamo simulation. 
		The Rayleigh number $Ra$
		in the simulation is given above the panels. }
	\label{ubdtdz}
\end{figure} 
\begin{figure}
	\centering
	\hspace{-2.3 in}	(a)  \hspace{1.8 in} (b) \\
	\includegraphics[width=0.33\linewidth]{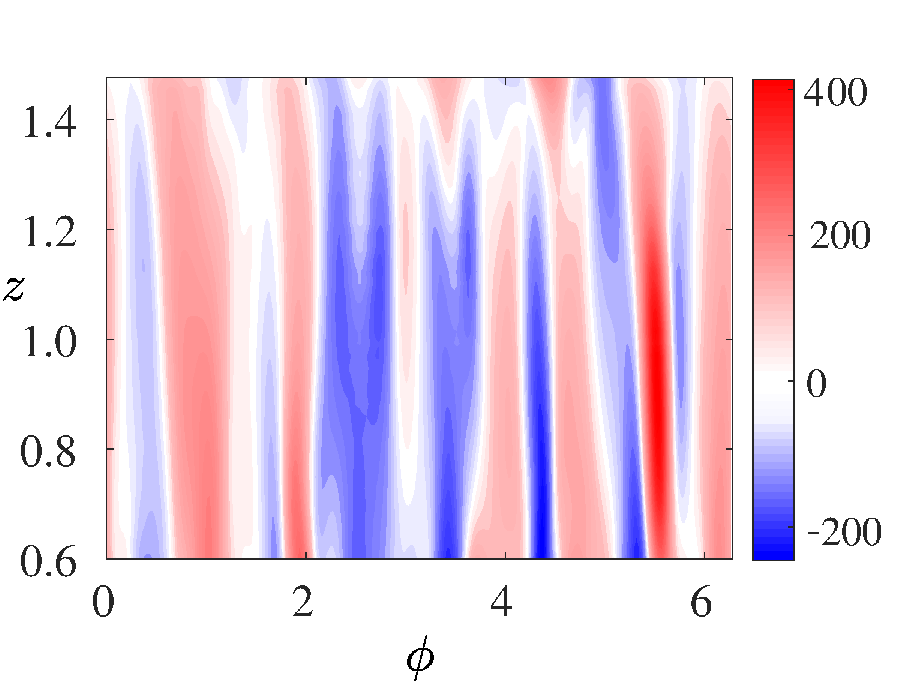}
	\includegraphics[width=0.33\linewidth]{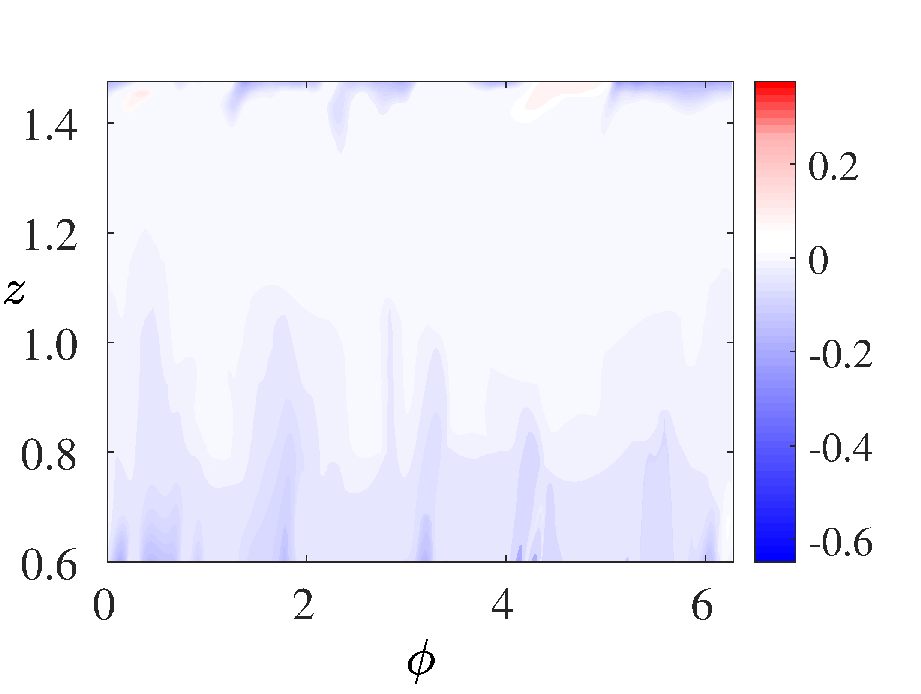}
	\caption{ Cylindrical section ($z-\phi$) plots 
		within the tangent 
		cylinder of $u_z$ (a) and $\partial T/\partial z$ (b)
		for the nonmagnetic convection simulation at
$E = 6\times 10^{-5}$, $Ra=12000$ and $Pr = 5$.}
	\label{ubnm}
\end{figure}
In Figure \ref{ubdtdz}, a cylindrical ($z$-$\phi$) section 
at selected cylindrical radii $s$ of the axial
field $B_z$, axial flow $u_z$ and axial temperature gradient
 $\partial T/\partial z$ are shown. 
The dynamo parameters are $E= 6\times 10^{-5}$ and $Pm=Pr=5$.
At $Ra=1000$, the magnetic flux that diffuses into the TC
from outside appears to concentrate
on length scales comparable to that at convective onset.
Because of the convergent flow at the base of an
upwelling \citep{gafd2006}, $B_z$ typically concentrates 
near the base of the TC. A plume 
appears at the same location as the flux concentration
(figures \ref{ubdtdz} (a) and (d)), and the same
behaviour is reproduced at much stronger forcing 
(figures \ref{ubdtdz} (b) and (e)). 
Outside the plume, where the magnetic flux is weak, convection is absent
except at the base of TC. This suppression of convection, 
also noted in linear
magnetoconvection with a laterally varying field \citep{jfm17a}, can be
explained by a localized unstable stratification produced
by the magnetic field through a magnetic-Coriolis balance 
(figures \ref{ubdtdz} (g) and (h)). Elsewhere,
the fluid layer is neutrally buoyant. There are stably stratified
regions with $\partial T/\partial z>0$  at the top of TC, where warm fluid carried
upward by the plume builds up. When buoyant forcing is increased
to the point of polarity reversals outside the TC, there is no
coherent magnetic flux concentration within
the TC (figure \ref{ubdtdz} (c)); then convection takes the form
of an ensemble of plumes (figure \ref{ubdtdz} (f)), 
similar to that found in nonmagnetic
simulations  (see figure \ref{ubnm} (a)).  In both nonmagnetic 
and reversing simulations, the entire fluid layer is 
unstably stratified (figures \ref{ubnm} (b) and \ref{ubdtdz} (i)).

\begin{figure}
	\centering
	\hspace{-0.8 in}	{\Large $B_z$ }  \hspace{1.75 in} {\Large $u_z$ } 
	  \hspace{1.75 in} {\Large $u_\phi$ }\\
	\hspace{-0.8 in}	$Ra=12000$\\
	\hspace{-2 in}	(a)  \hspace{1.75 in} (b)  \hspace{1.75 in} (c) \\
	\includegraphics[width=0.3\linewidth]{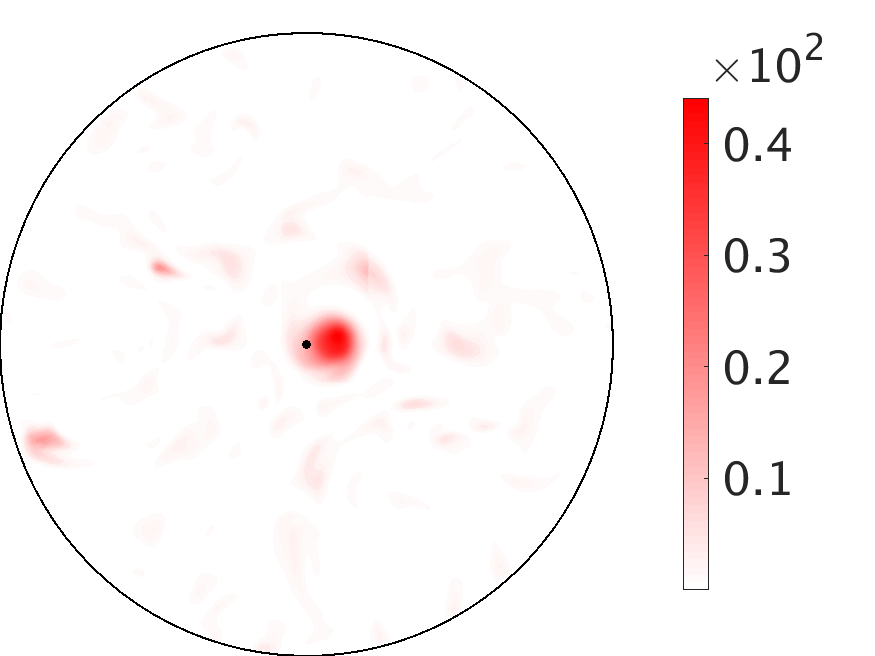}
	\includegraphics[width=0.3\linewidth]{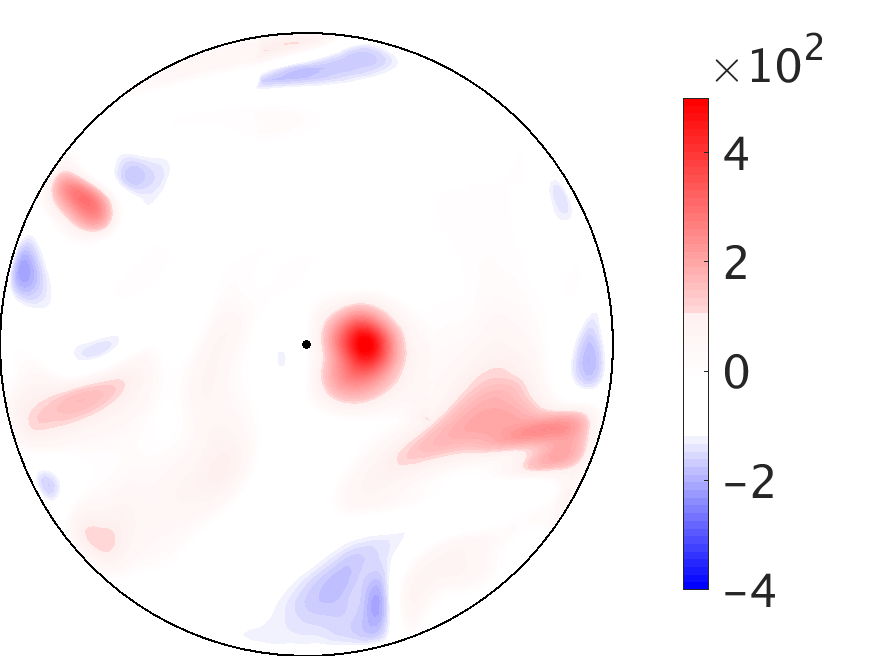}
	\includegraphics[width=0.3\linewidth]{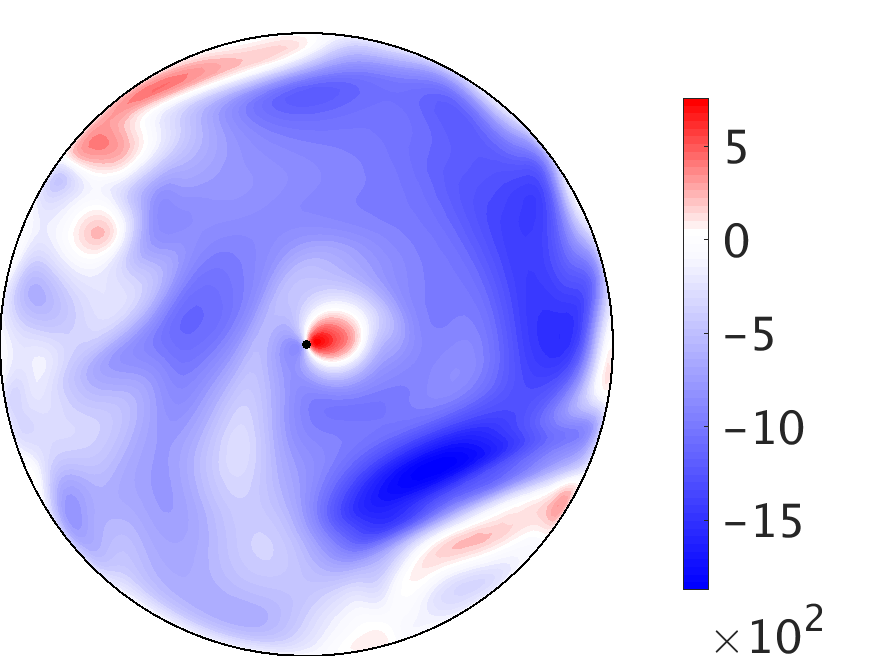}\\
	\hspace{-0.8 in}	$Ra=18000$\\
	\hspace{-2 in}	(d)  \hspace{1.75 in} (e)  \hspace{1.75 in} (f) \\
	\includegraphics[width=0.3\linewidth]{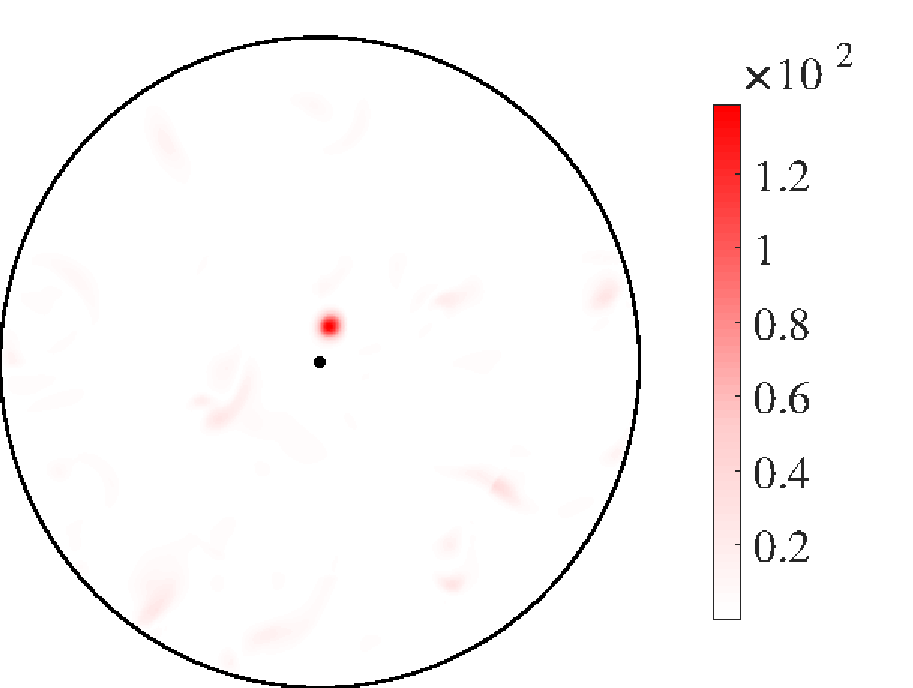}
	\includegraphics[width=0.3\linewidth]{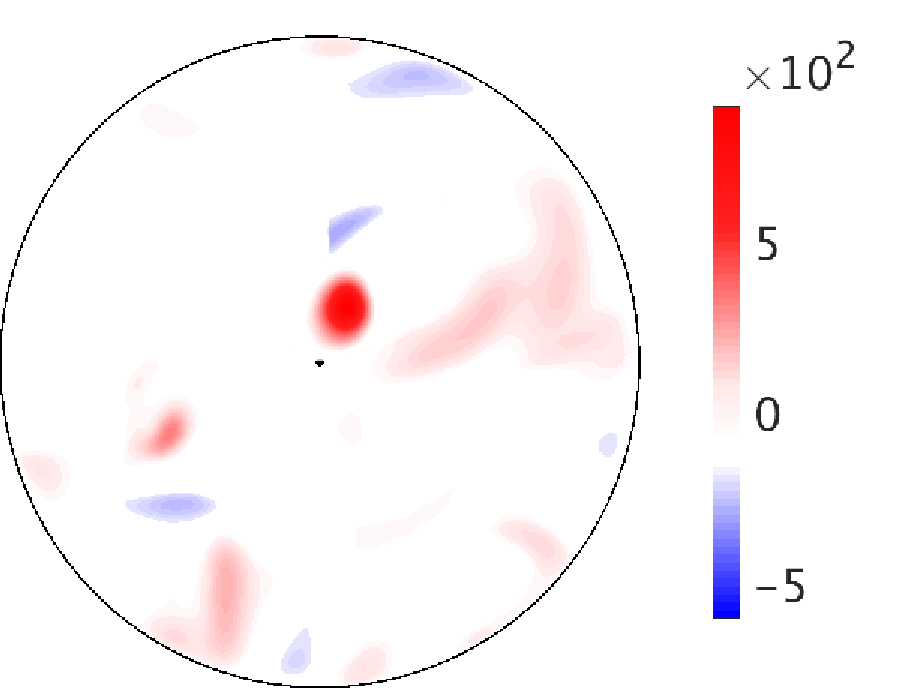}
	\includegraphics[width=0.3\linewidth]{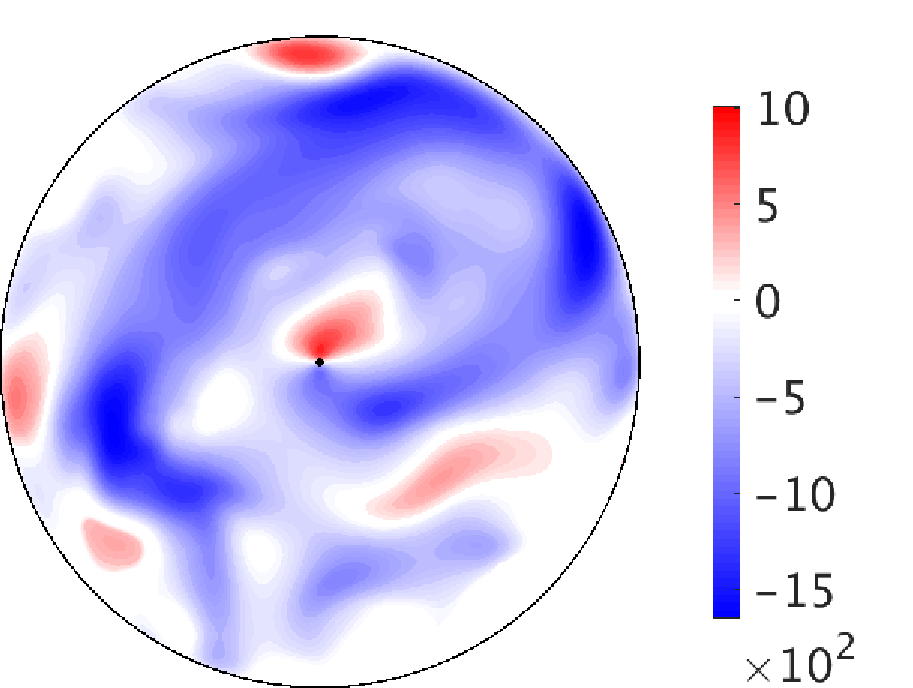}\\
	\hspace{-0.8 in}	$Ra=21000$\\
	\hspace{-2 in}	(g)  \hspace{1.75 in} (h)  \hspace{1.75 in} (i) \\
	\includegraphics[width=0.3\linewidth]{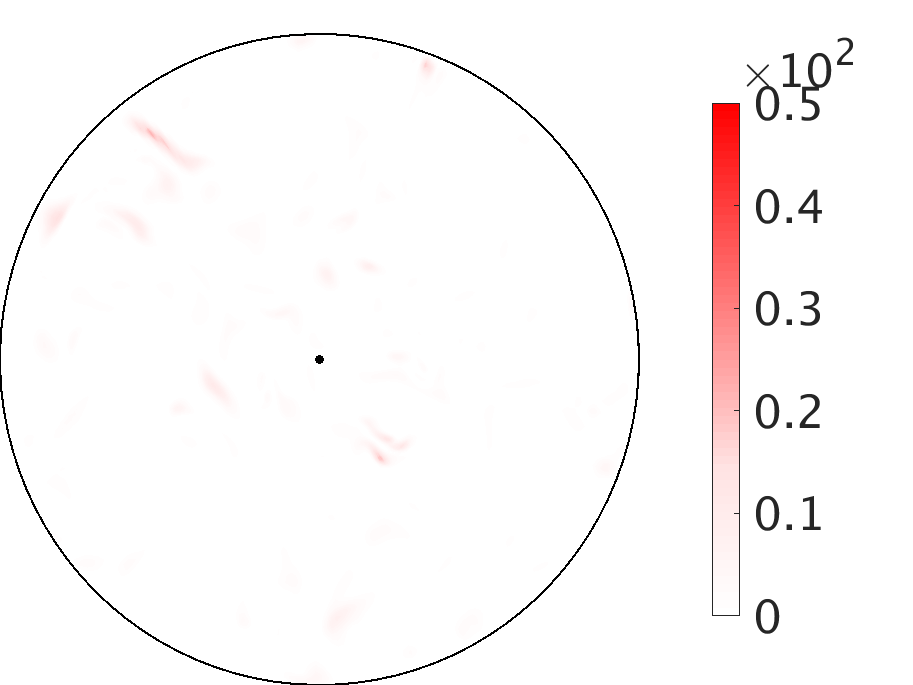}
	\includegraphics[width=0.3\linewidth]{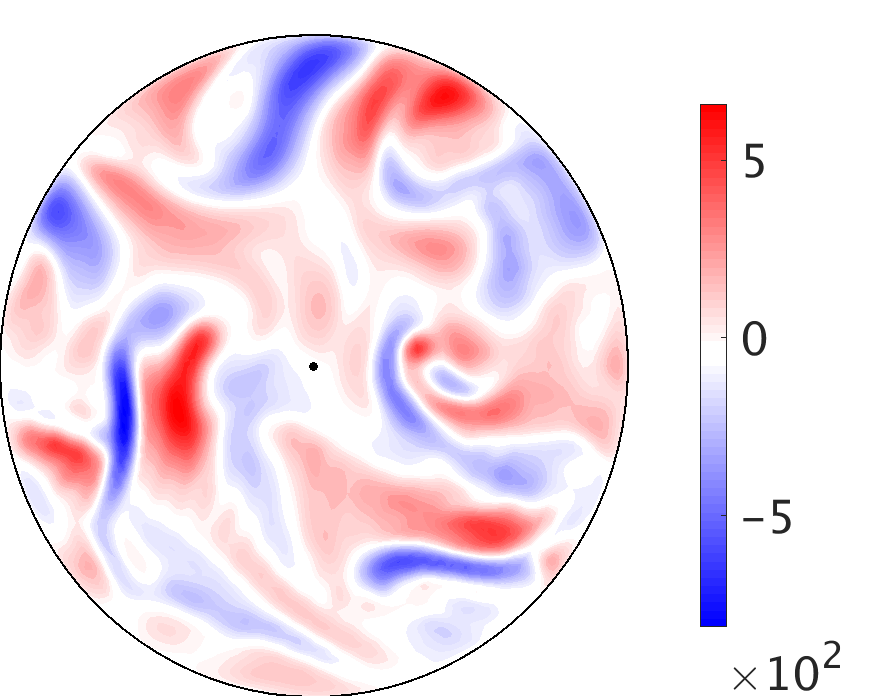}
	\includegraphics[width=0.3\linewidth]{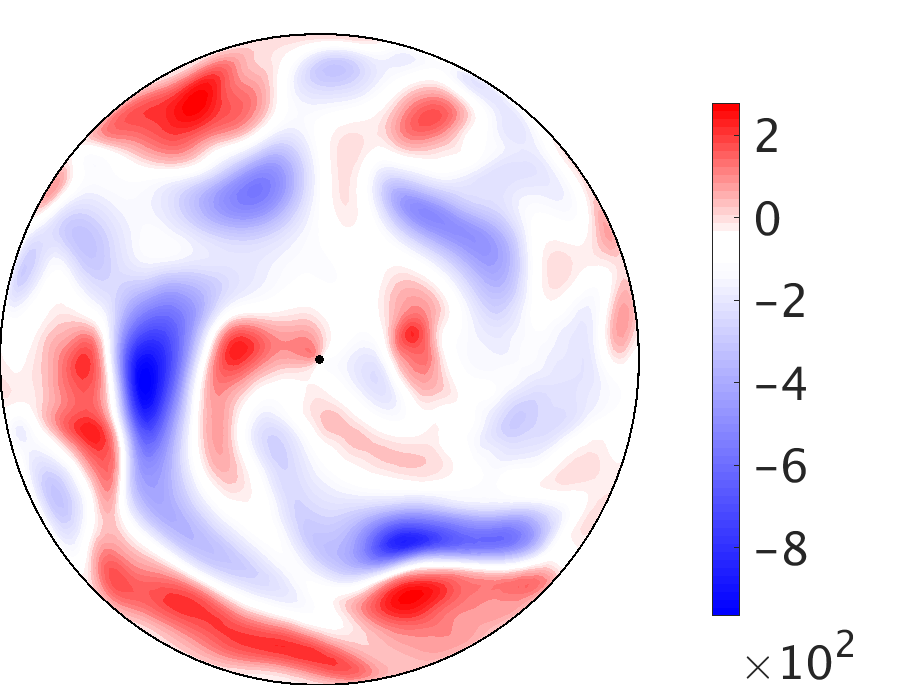}\\
	\caption{Horizontal $(z)$ section plots within the tangent cylinder 
		of the axial magnetic field $B_z$  at $z=0.9$ (left panels), $u_z$  
		at $z=1.4$ (middle panels) and $u_\phi$  at $z=1.4$ (right
panels) for 
	$Ra=12000$ (a-c), $Ra=18000$ (d-f), $Ra=21000$ (g-i). 
		The other dynamo parameters are 
$E=6 \times 10^{-5}, Pm=Pr=5$. The figures are shown 
from the saturated state of the dynamo simulation.}
	\label{ubphi}
\end{figure}

\begin{figure}
	\centering
	\hspace{-1.9in} (a) \hspace{1.8 in}(b) \\
	\includegraphics[width=0.3\linewidth]{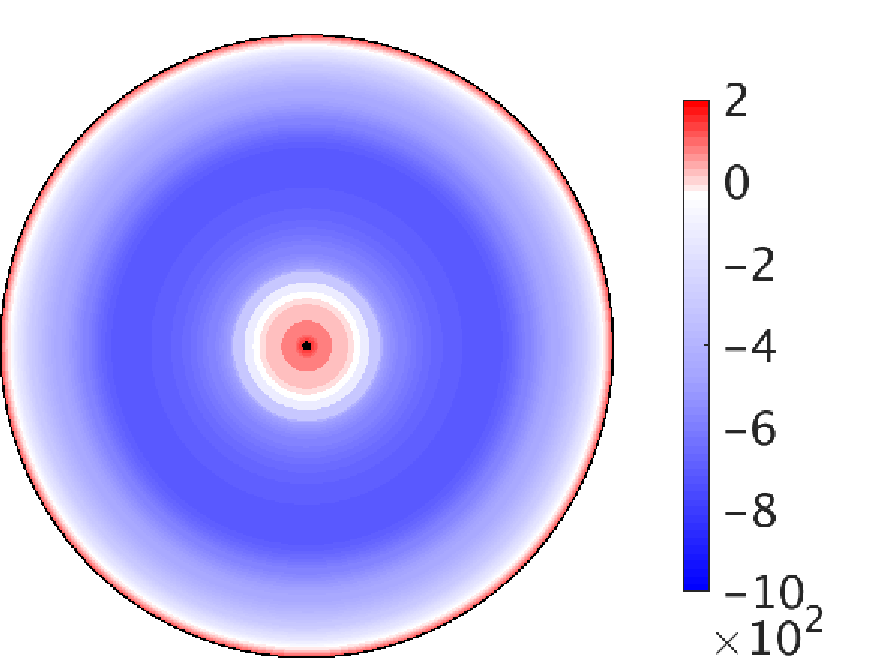}
	\includegraphics[width=0.3\linewidth]{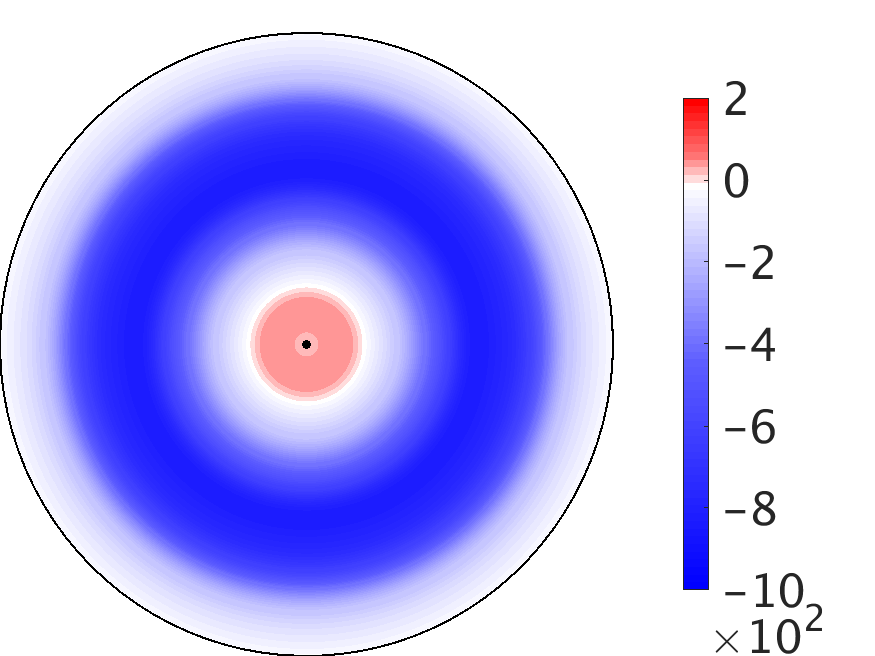}\\
	\hspace{-1.9in} (c) \hspace{1.8 in}(d) \\
	\includegraphics[width=0.3\linewidth]{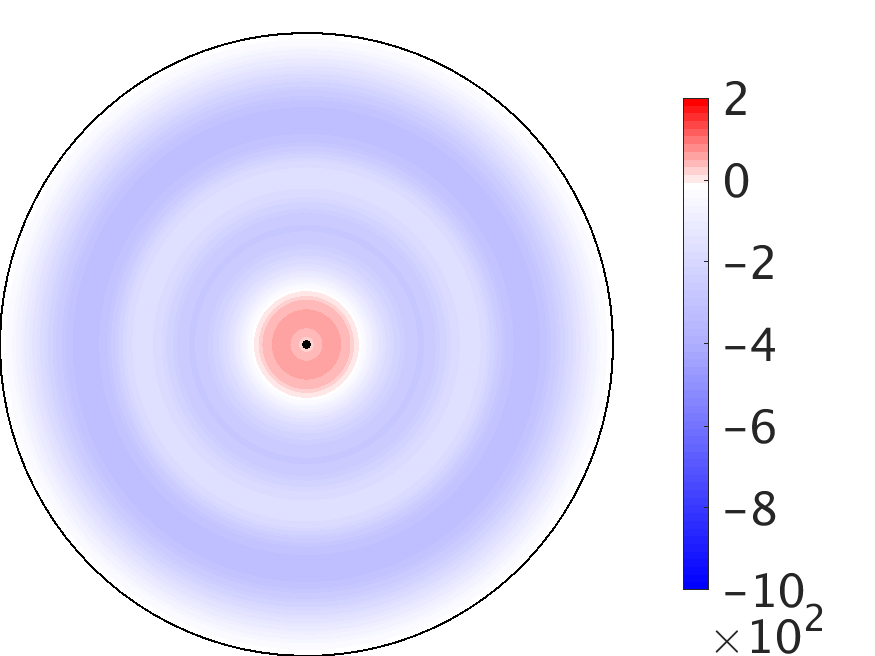}
	\includegraphics[width=0.3\linewidth]{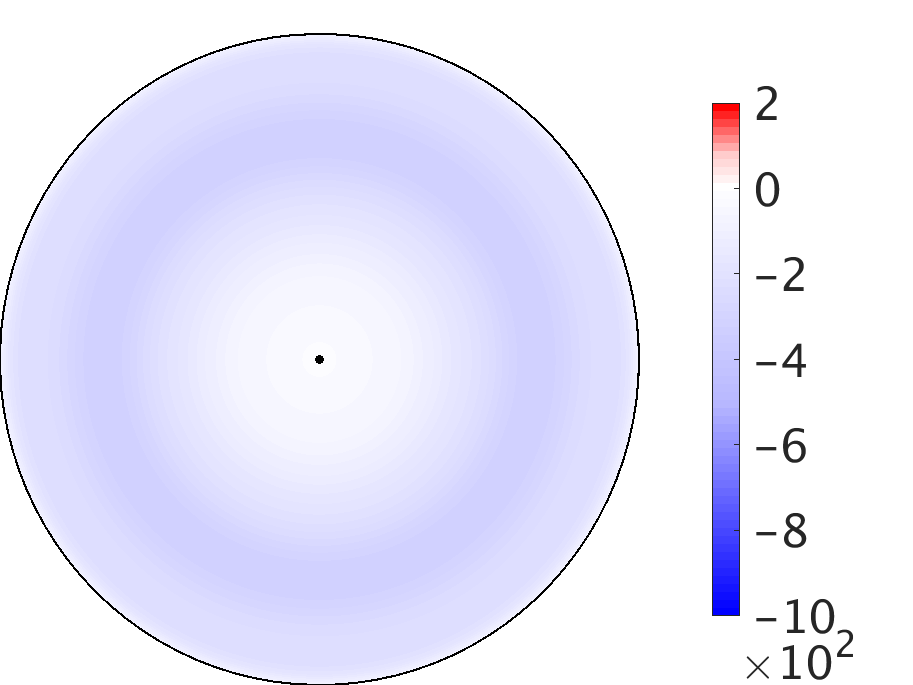}\\
	\caption{Horizontal $(z)$ section plots at height $z=1.4$ above the
		equator of the time and azimuthally 
		averaged flow $u_\phi$ within the TC
		at $E=6\times10^{-5}$, $Pm=Pr=5$. 
		(a) $Ra=$12000, (b) $Ra=$18000, (c) $Ra=$21000, 
		(d) nonmagnetic run at $Ra=$12000. 
		All figures represent the saturated 
		state of the simulations.}
	\label{pvavg}
\end{figure}

The polar vortices produced by the pattern of convection 
within the TC are shown in the $z$-section
plots in figures \ref{ubphi}. For a wide
range of $Ra$ in the dipole-dominated regime, the correlation
between $u_z$ and $B_z$ exists (figures 
\ref{ubphi} (a), (b) and (d), (e)); the radially outward motion at
the top of the plume is turned into a strong anticyclonic vortex
by the Coriolis force (figures \ref{ubphi} (c), (f)). 
In the strongly driven regime of dipole collapse, the absence of a coherent
field results in multi-column convection
(figure \ref{ubphi} (g), (h)) and a much weaker
anticyclonic circulation near the poles (figure \ref{ubphi} (i)).
On time and azimuthal average, the polar vortex intensity 
increases with $Ra$ (figures \ref{pvavg} (a) and (b)); 
however, at the point of reversals, the vortex
intensity significantly diminishes (figure \ref{pvavg} (c)), resembling that
in the nonmagnetic runs (figure \ref{pvavg} (d)). 

The maximum dimensionless time and azimuthally averaged value of 
$u_\phi$ is measured within the TC, with its radial distance
from the rotation axis. 
For example, for $Ra=12000$ ($E=6\times 10^{-5}, Pm=Pr=5$), 
this magnitude of $u_\phi$ 
is 508, at radius 0.33. This could be scaled up to its value 
in the Earth's core \citep{grl2005}, giving
\begin{equation}\label{scuphi}
	\begin{aligned}
		u_{\phi,\mbox{sc}}=\frac{u_\phi \eta}{L}
		=2.2478\times 10^{-4} 
		\mbox{m} \mbox{s}^{-1} 
		\approx 0.535 ^\circ \mbox{yr}^{-1},
	\end{aligned}
\end{equation}

where $\eta$ and $L$ have the values 1 m$^2$s$^{-1}$ 
and $2.26\times10^6$ m respectively. The values of 
$u_{\phi,\mbox{sc}}$
in the simulations and their respective values in 
nonmagnetic simulations
are given in  table \ref{parameters}.
Increasing the strength of forcing in the nonmagnetic
simulations does not result in stronger polar vortices,
 which indicates the crucial role of the magnetic field
in generating strong polar circulation.

\subsection{Magnetic waves in the TC}
\label{magtc}

\begin{figure}
	\centering
	\hspace{-2.5 in}	(a)  \hspace{2.75 in} (b) \\
	\includegraphics[width=0.45\linewidth]{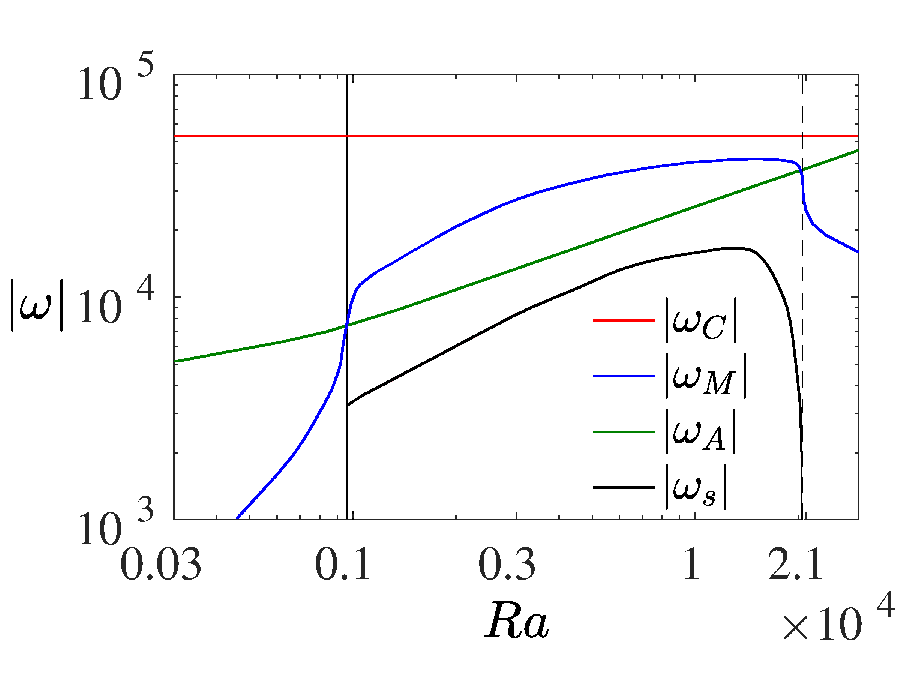}
	\includegraphics[width=0.45\linewidth]{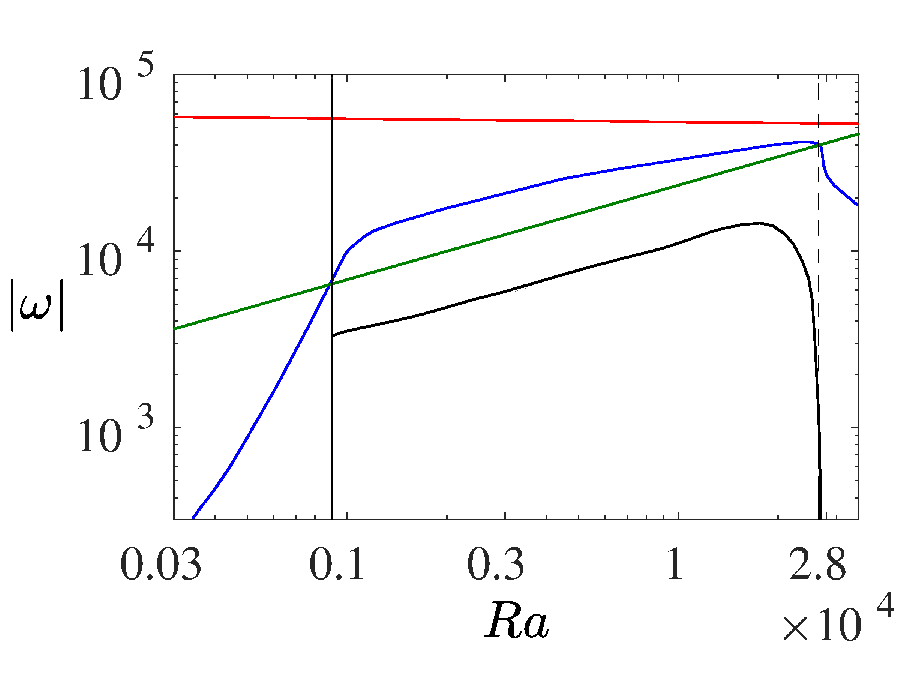}\\
	\caption{(a, b) Absolute values of the measured frequencies $\omega_C$, 
		$\omega_M$, $\omega_A$, and $\omega_s$ 
inside the tangent cylinder versus the Rayleigh number $Ra$
in the saturated dynamo. The dynamo
parameters are (a) $E=6\times 10^{-5}$, $Pm=Pr=5$ and  
	(b) $E=1.2\times 10^{-5}$, $Pm=Pr=1$. The solid vertical lines 
		indicate the onset of the slow MAC
waves inside the tangent cylinder while the dashed vertical lines mark the 
suppression of the slow waves. 
The onset of the slow waves
occurs at $Ra=960$ (a) and $Ra=900$ (b) and their suppression
occurs at $Ra=20500$ (a) and $Ra=27500$ (b).}
	\label{freqtc}
\end{figure}

Isolated density disturbances within the TC  evolve 
as fast and slow MAC waves in the presence of the rapid
rotation and magnetic field. Since the frequency of these waves depends 
on the fundamental frequencies $\omega_C$, $\omega_M$ and $\omega_A$
(Section \ref{problem_setup}), we look at their magnitudes
in dimensionless form \citep{varma2022},

\begin{subequations}\label{dimlessfreq}    
	\begin{gather}  
		\omega_C^2=\frac{Pm^2}{E^2}\frac{k_z^2}{k^2},		
		\quad 	\omega_M^2=\frac{Pm}{E}(\bm{B}\cdot \bm{k})^2,	
		\quad	 -\omega_A^2=\frac{Pm^2 Ra}{Pr E}\frac{k_h^2}{k^2},\tag{\theequation a-c} 
	\end{gather}
\end{subequations}

where the Alfv\'en frequency is based on the three components of
the magnetic field at the peak field location
and the frequencies are scaled by $\eta/L^2$. Here, $k_s$, 
$k_\phi$ and $k_z$ are the radial, azimuthal and 
axial wavenumbers in cylindrical coordinates, $k_h$ is the 
horizontal wavenumber inside the TC given by $k_h^2=k_s^2+k_\phi^2$, 
and $k^2=k_s^2+k_\phi^2+k_z^2$.
The wavenumbers are calculated within the TC as the focus 
of this study is to investigate the role of 
MAC waves inside the TC. 
In figure \ref{freqtc}, the dimensionless
frequencies in the saturated dynamo are calculated using the 
mean values of the wavenumbers. 
For example, real space
integration over $(s,\phi)$ gives 
the kinetic energy as a function of $z$, the
Fourier transform of which gives the 
one-dimensional spectrum $\hat{u}^2(k_z)$.
In turn, we obtain,
\begin{equation}\label{eq:kz}
	\begin{aligned}
		\bar{k}_z=\frac{\sum k_z 
		\hat{u}^2(k_z)}{\sum \hat{u}^2(k_z)}.
	\end{aligned}
\end{equation}
A similar approach gives $\bar{k}_s$ and $\bar{k}_\phi$. Since the 
flow length scale transverse to the rotation axis is comparable to that
at convective onset within the TC, the mean 
wavenumbers are calculated over the entire
spectrum without scale separation.

As $Ra$ is increased progressively, $|\omega_M|$ exceeds $|\omega_A|$,
which is when an isolated plume forms within the TC. The inequality
$|\omega_C| > |\omega_M| > |\omega_A| > |\omega_\eta|$ exists for
wide range of $Ra$ until the point of polarity transitions,
indicating the active presence of MAC waves within the TC
(figure \ref{freqtc} (a) \& (b)). 
Further
increase of forcing results in the state marked
by the dashed vertical lines where 
$|\omega_A| \approx |\omega_M|$, at which
the slow wave frequency $\omega_s$ goes to zero. 
As the slow waves are suppressed, the flow within the TC
takes the form of an ensemble of plumes, similar
to that in nonmagnetic convection.
Since fast waves of frequency
$\omega \sim \omega_C$ exist in both magnetic and nonmagnetic
convection, the
excitation of slow MAC waves
may be crucial in the formation of isolated off-axis plumes, which
in turn produce strong anticyclonic polar vortices. We pursue
this idea by measuring wave motions within the TC in
saturated dynamos.
\begin{figure}
	\centering
	\hspace{-3 in}	(a)  \hspace{3 in} (b) \\
	\includegraphics[width=0.45\linewidth]{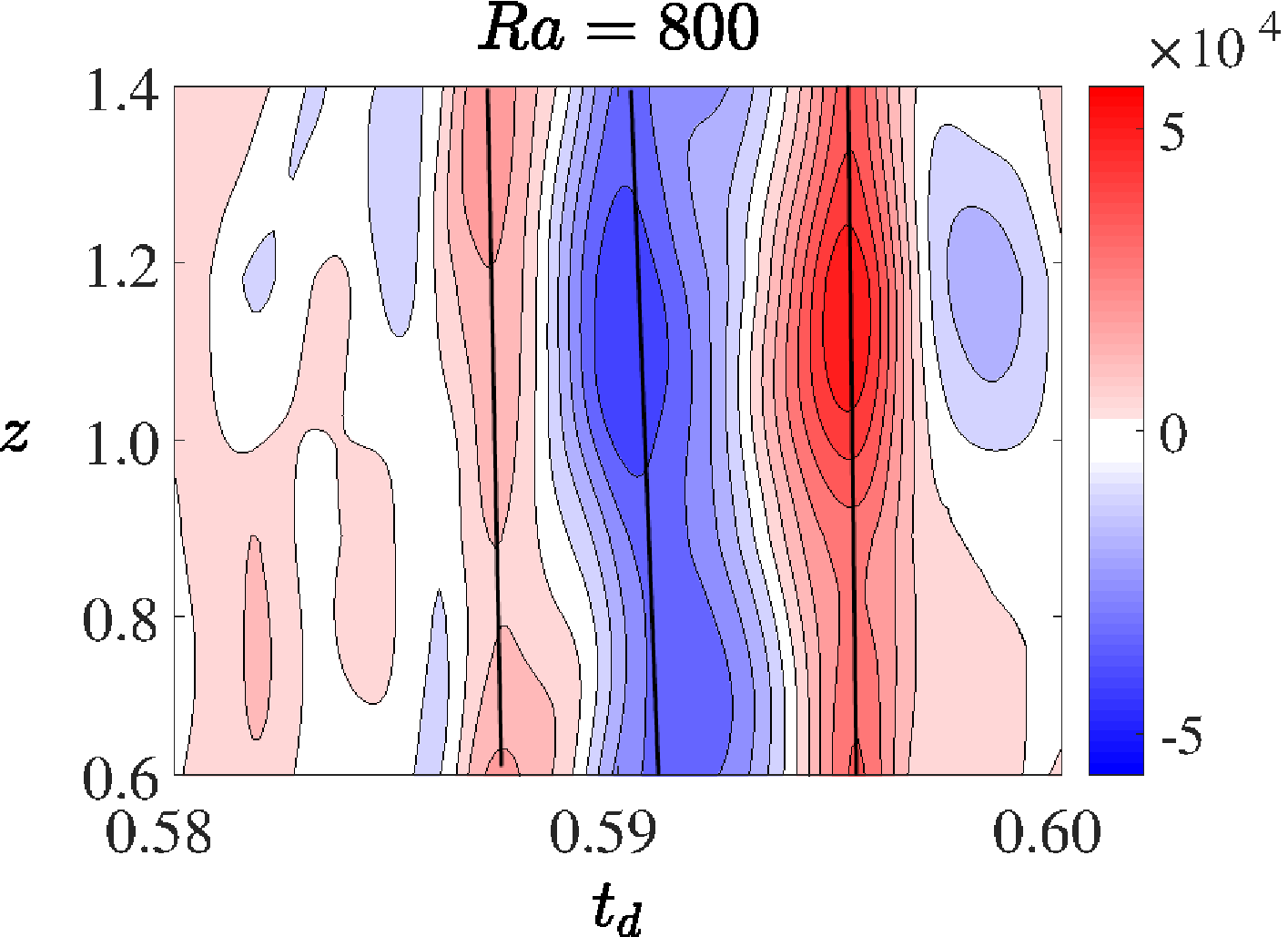}
	\includegraphics[width=0.45\linewidth]{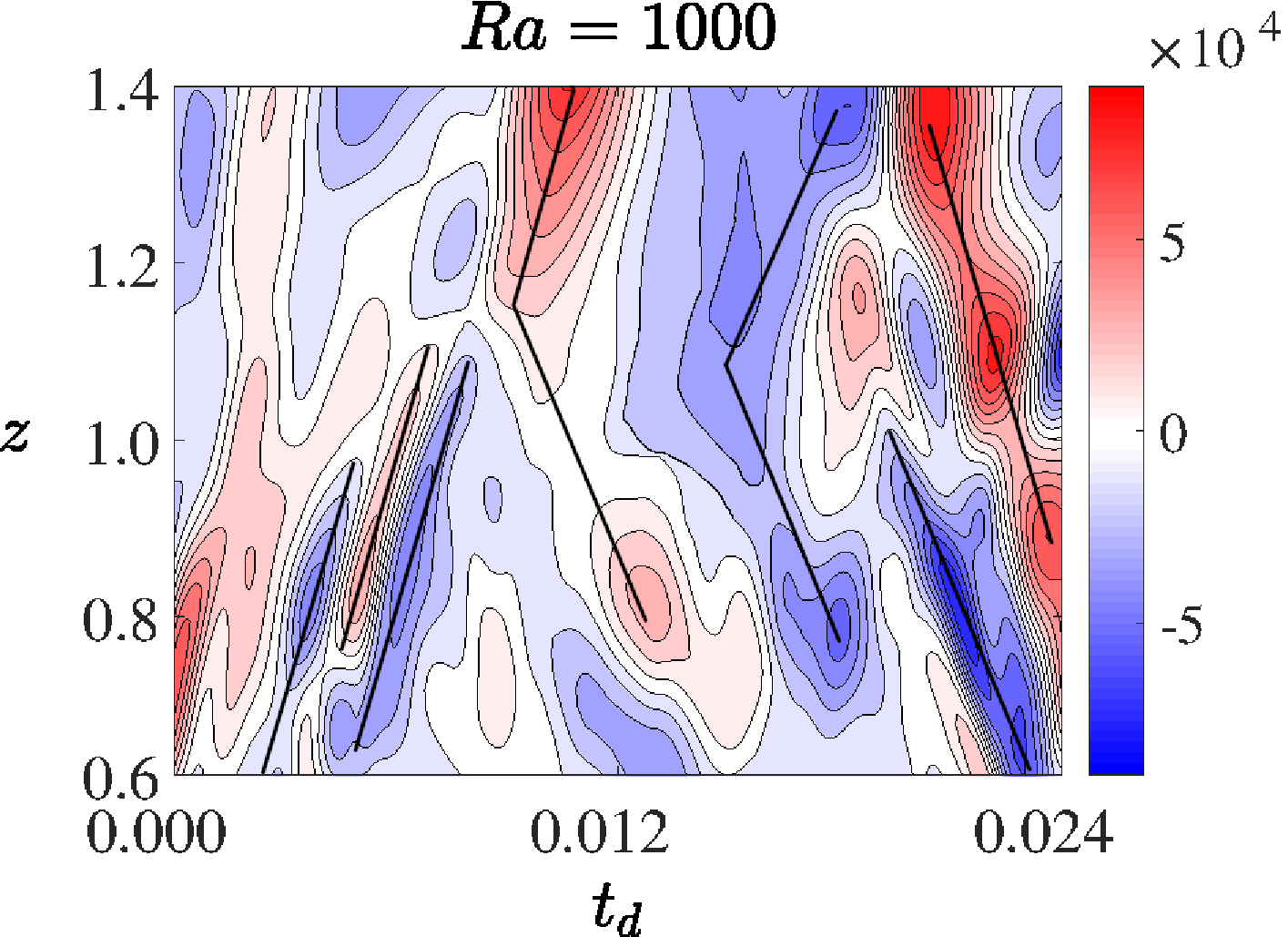}\\
	\hspace{-3 in}	(c)  \hspace{3 in} (d) \\
	\includegraphics[width=0.45\linewidth]{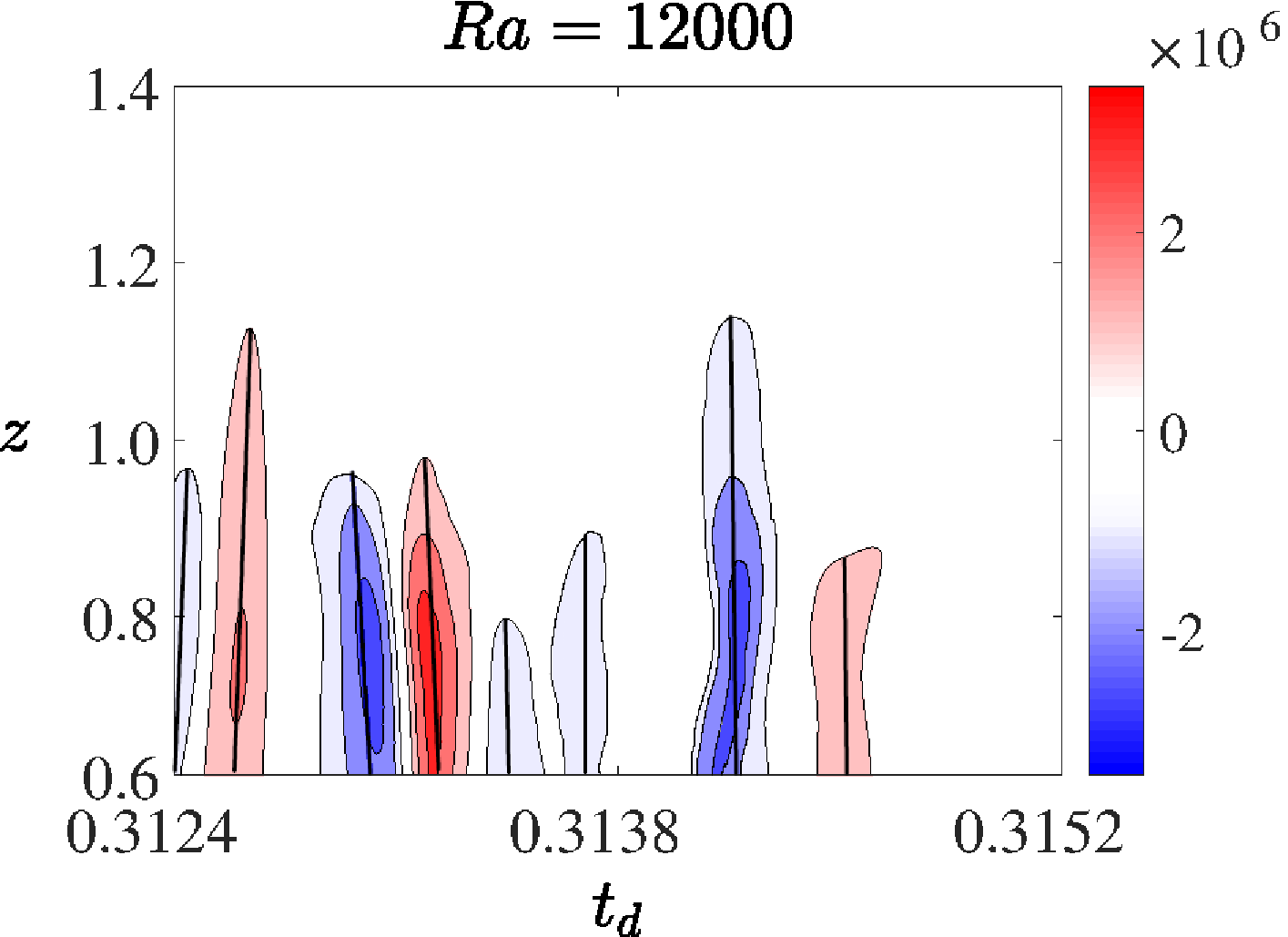}
	\includegraphics[width=0.45\linewidth]{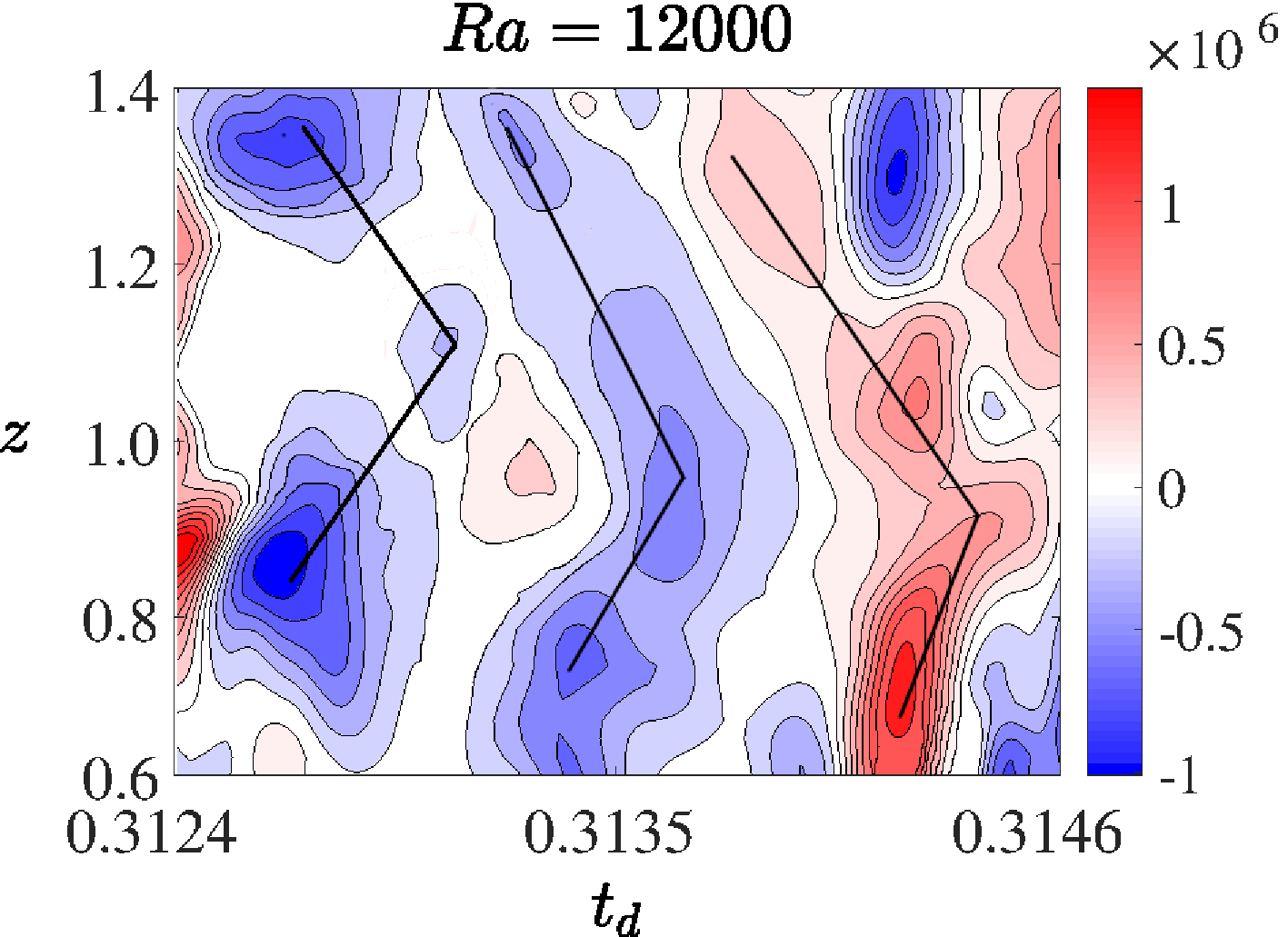}\\
	\hspace{-3 in}	(e)  \hspace{3 in} (f) \\
	\includegraphics[width=0.45\linewidth]{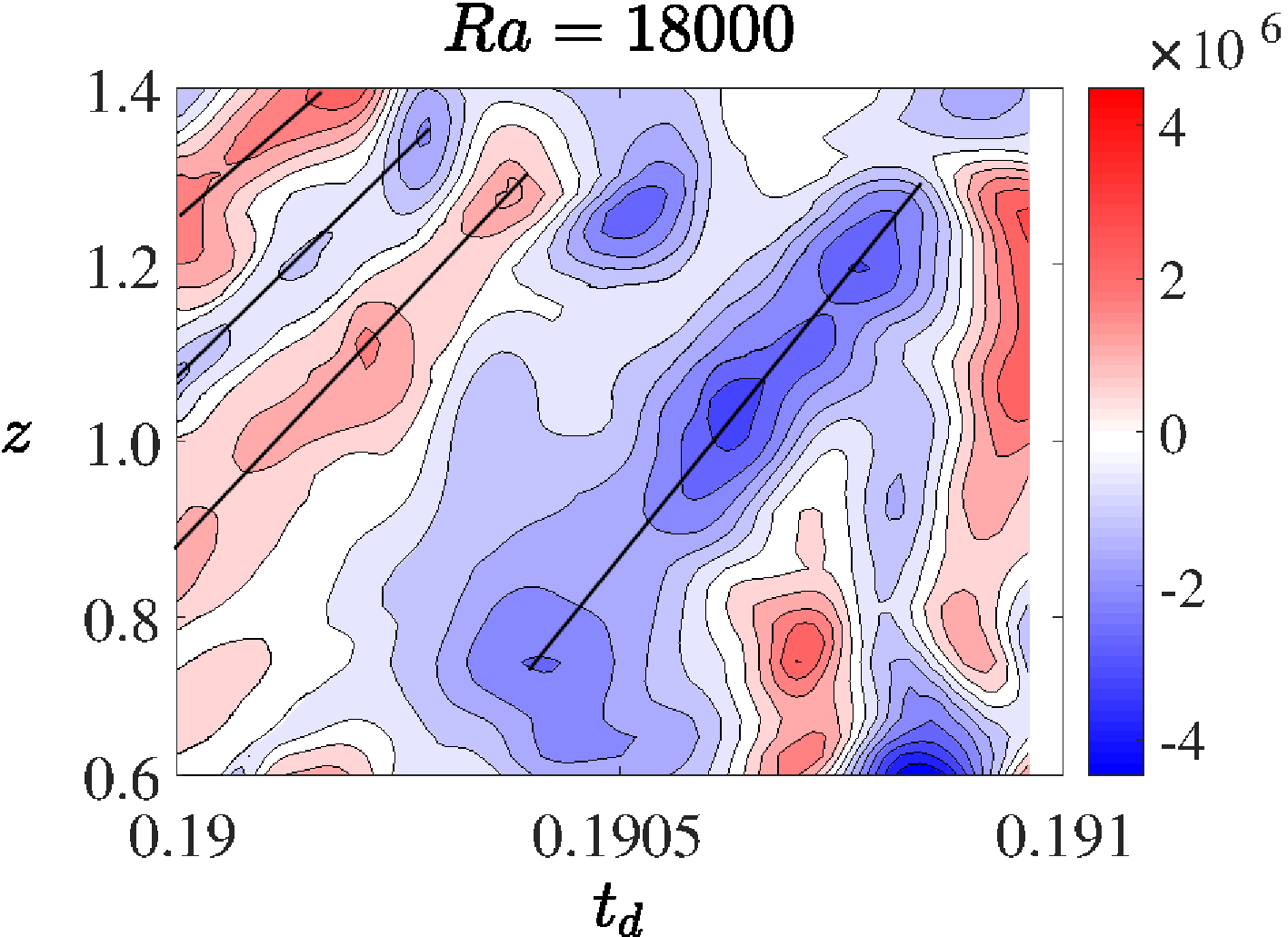}
	\includegraphics[width=0.45\linewidth]{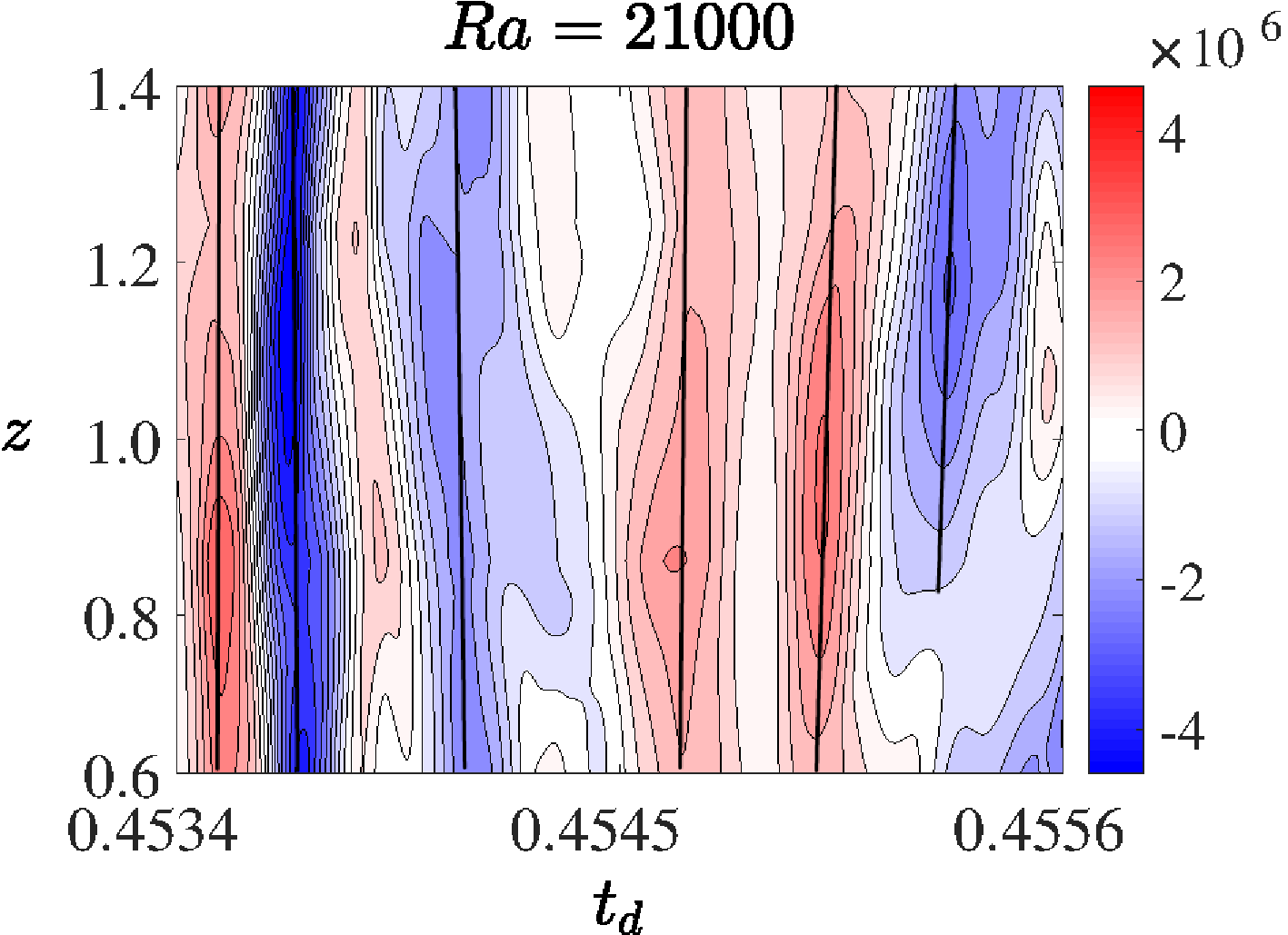}\\
	
	\caption{Contour plots of $\partial u_z/\partial t$ at 
		cylindrical radius $s = 0.3$ for small intervals of 
		time in the saturated state. The parallel black 
		lines indicate the predominant direction of travel 
		of the waves and their slope gives the group velocity.
		The Rayleigh number $Ra$ of the simulation is 
		given above each panel. The other dynamo parameters 
		are $E = 6\times10^{-5}, Pr = Pm = 5$. The estimated 
		group velocity of the fast and slow MAC waves ($U_f$ 
		and $U_s$ respectively) and the measured group velocity 
		$U_{g,z}$ are given in table \ref{cgvalues}.}
	\label{cg}
\end{figure}
In figure \ref{cg} , 
contours of $\partial u_z/\partial t$ are plotted for small time windows of
 approximately constant ambient
 magnetic field and wavenumbers at cylindrical radius $s=0.3$. 
The axial motion is measured by considering
the full spectrum. In
table \ref{cgvalues}, the measured axial 
group velocity of the waves, $U_{g,z}$
is compared
with the estimated group velocity of the fast and slow MAC waves,
given by $U_f$ and $U_s$ respectively. 
At $Ra=800$, convection is initiated within the TC under a 
weak ambient magnetic 
field. Consequently, only 
fast MAC waves are excited within the TC (figure \ref{cg}(a)). 
At $Ra = 1000$, the onset of slow MAC waves occurs
(figure \ref{cg}(b)); however, 
convection near the base of the TC is predominantly 
produced by the excitation of fast MAC waves 
(figure \ref{cg} (c)).
At the location $(s,\phi)$ of the plume, the 
measured group
velocity (figures \ref{cg} (b), (d) and (e)) 
matches well with that
of the slow MAC waves. At the onset of 
polarity transitions,
 the slow waves disappear, and the 
 homogeneous
convection within the TC is entirely made up of
fast waves (figure \ref{cg} (f)).

\begin{figure}
	\centering
	\hspace{-3 in}	(a)  \hspace{3 in} (b) \\
	\includegraphics[width=0.45\linewidth]{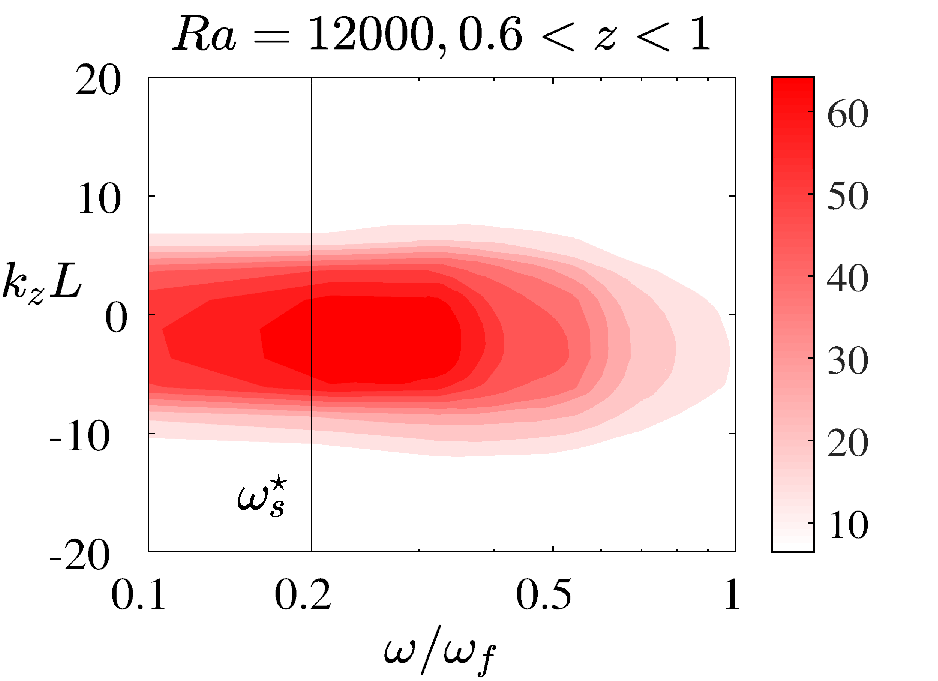}
	\includegraphics[width=0.45\linewidth]{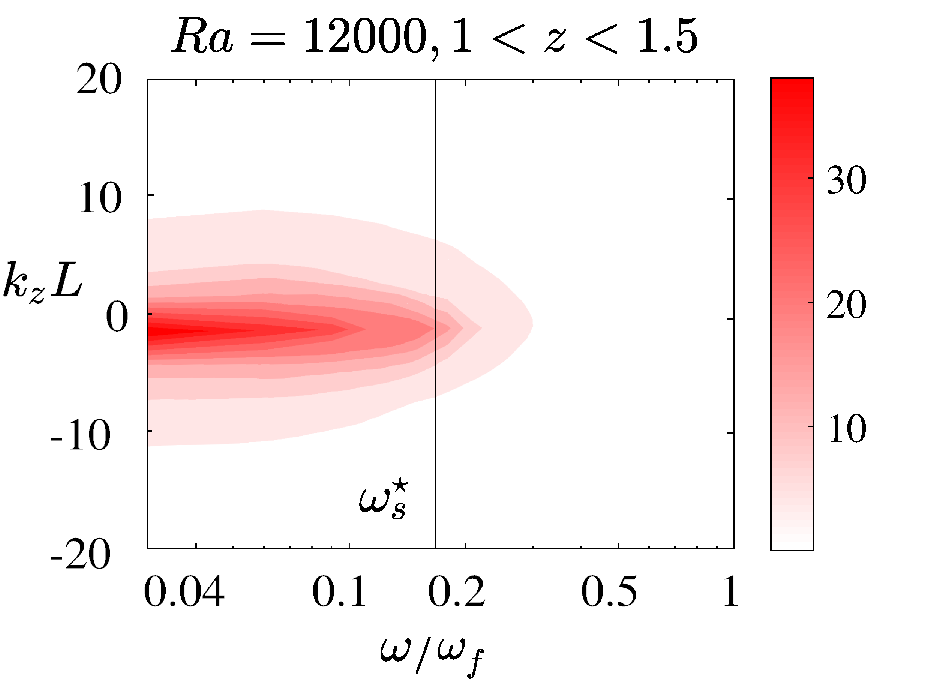}\\
	\hspace{-3 in}	(c)  \hspace{3 in} (d) \\
	\includegraphics[width=0.45\linewidth]{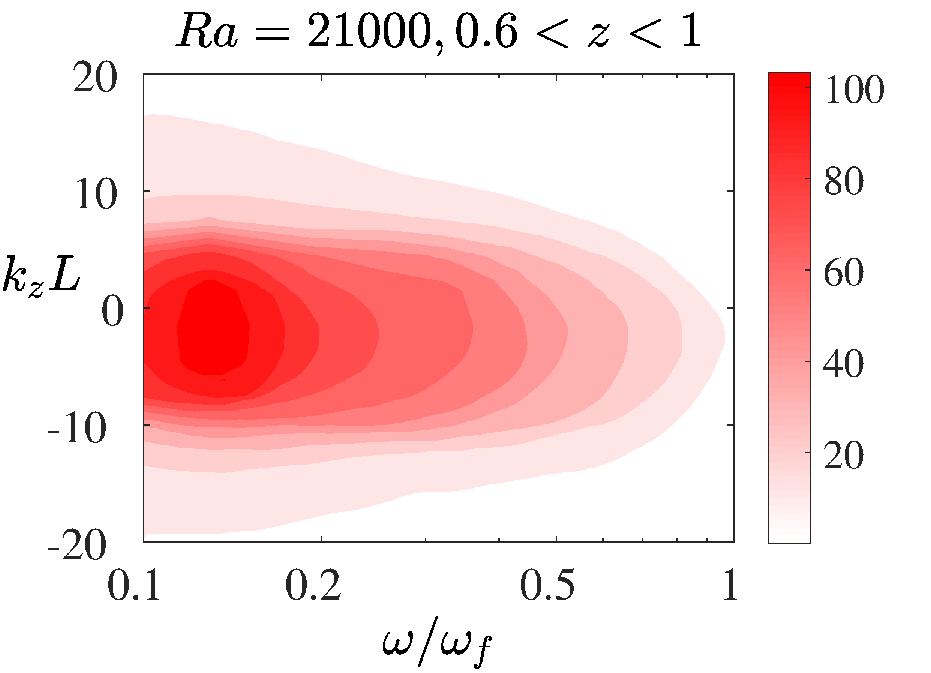}
	\includegraphics[width=0.45\linewidth]{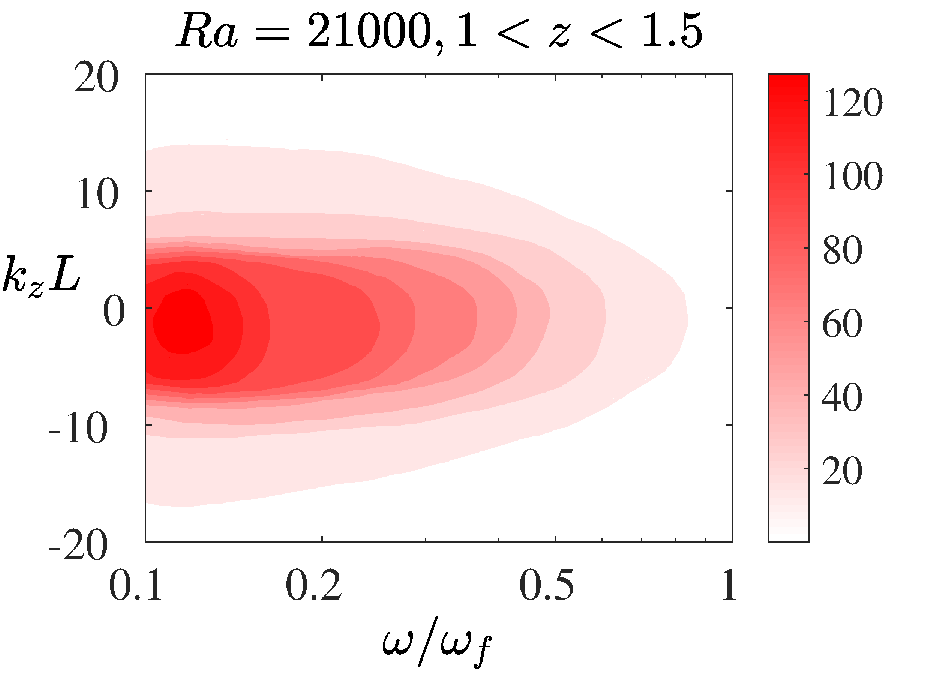}\\
	
	\caption{FFT spectra of $\partial u_z/\partial t$ for $0.6<z<1$ 
		and $1<z<1.5$ inside tangent cylinder. The spectra are 
		computed at discrete $\phi$ points and then averaged 
		azimuthally from the saturated states of the dynamo runs. 
		The thin vertical lines 
		in (a) and (b) give $\omega_s^\star = \omega_s/\omega_f$,
		 where $\omega_f$ and $\omega_s$ are 
		 the estimated fast and 
		 slow MAC wave frequencies. 
		 The Rayleigh number in the 
		 simulation is given above each panel. 
	The other dynamo parameters are 
	$E = 6\times10^{-5}$ and $Pm = Pr =5$.}
	\label{fft}
\end{figure}

Figure \ref{fft} shows the fast Fourier transform (FFT) of
$\partial u_z/\partial t$ separately for the 
lower ($0.6<z<1$) and upper ($1<z<1.5$) regions of the TC. 
The spectra were computed at discrete $\phi$ points 
and then azimuthally averaged. 
The thin vertical lines in figures \ref{fft} 
(a) and (b) give the values of $\omega_s^\star= \omega_s/\omega_f$, 
where $\omega_f$ and $\omega_s$ are 
the estimated 
frequencies of the fast and slow MAC waves. 
In the lower region of the TC, where the fast and slow 
waves coexist, the
 spectra suggest the dominance of the fast waves. 
However, the slow waves are dominant in the upper region of the TC. 
In the regime of polarity transitions ($Ra=21000$), 
the fast waves are dominant throughout the TC.

\begin{figure}
	\centering
	\hspace{-2.9 in}	(a)  \hspace{2.7 in} (b) \\
	\includegraphics[width=0.45\linewidth]{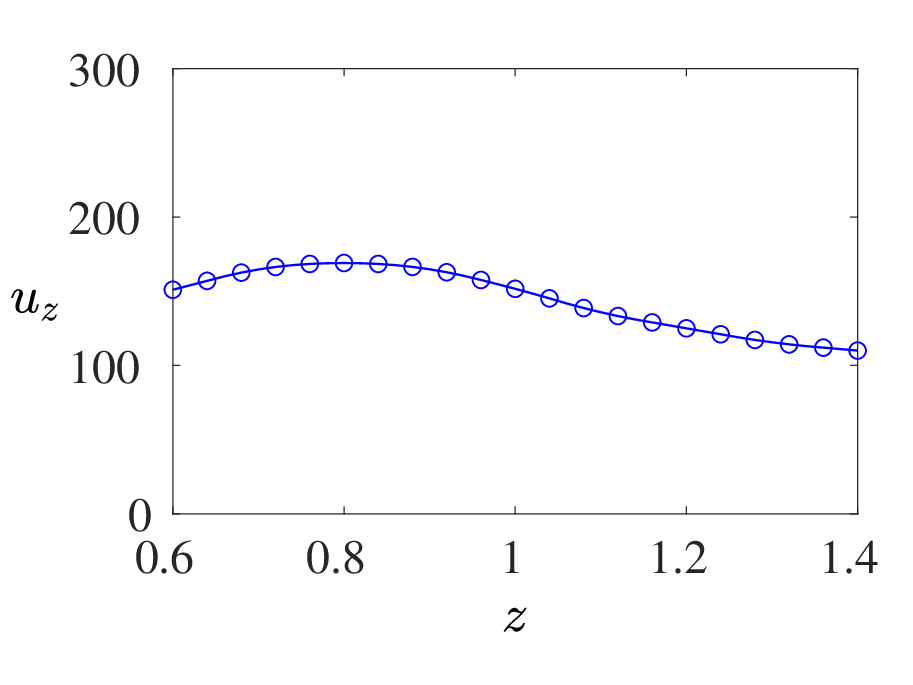}
	\includegraphics[width=0.45\linewidth]{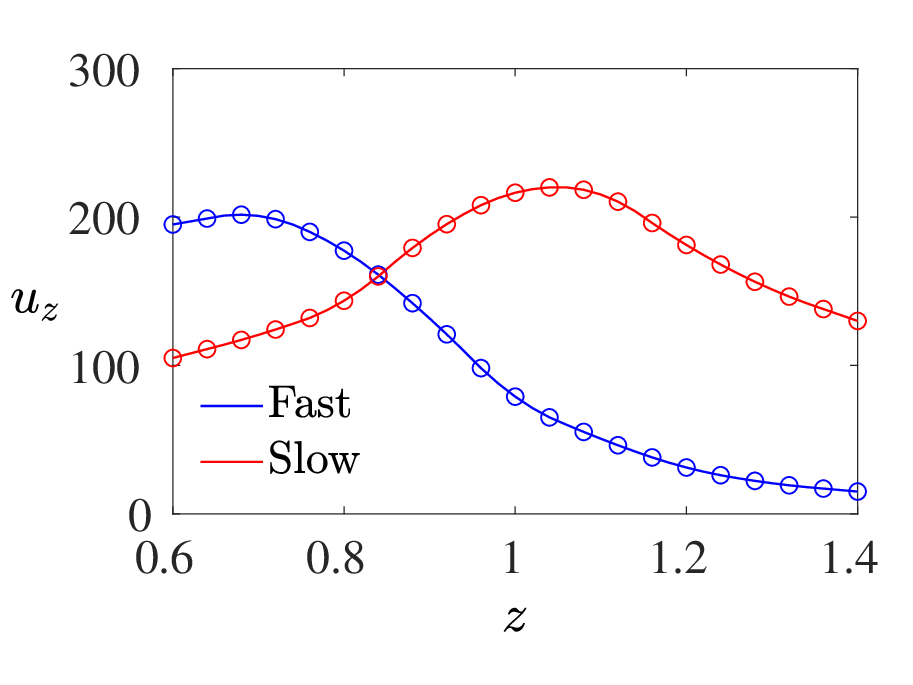}
	\caption{  Variation of the 
	time-averaged peak value of the axial ($z$)
 velocity with $z$ within the TC for (a) $Ra=800$ and (b) $Ra=1000$.
		The other dynamo parameters are 
		$E=6 \times 10^{-5}, Pm=Pr=5$.
}
	\label{onset}
\end{figure}

Having understood from figures \ref{cg}
and \ref{fft} that convection within the TC
is made up of fast and slow MAC waves, we examine
the magnitude of the wave motions in the neighbourhood
of convective onset.
In figure \ref{onset}, the magnitude of
the time-averaged 
peak $z$ velocity within the TC is plotted with respect to
 $z$ in two dynamo simulations near onset.
  At $Ra=800$, convection onsets via fast MAC waves, whose
  velocity does not change appreciably with $z$. At $Ra = 1000$,
  the axial magnetic field concentrates within the TC, giving
  $|\omega_M/\omega_C| \approx 0.2$ (table \ref{parameters}).
  Here, 
  the fast MAC wave velocity decreases 
  appreciably in the region $z>1$,
  where the slow waves propagate as isolated plumes.
Interestingly, the peak intensities of the fast and slow
wave motions are approximately equal (see also 
table \ref{comp}). The measurement of
wave motions in the dynamo model enables
meaningful comparisons of the dynamics within the TC
with that predicted by linear magnetoconvection.

\begingroup
\setlength{\tabcolsep}{2.7pt} 
\renewcommand{\arraystretch}{0.7} 

\begin{table}
\centering
\begin{tabular}{c@{\hspace{6pt}}c@{\hspace{6pt}}c@{\hspace{6pt}}c@{\hspace{6pt}}c@{\hspace{6pt}}c@{\hspace{7pt}}c@{\hspace{7pt}}c@{\hspace{7pt}}c@{\hspace{6pt}}c@{\hspace{7pt}}c@{\hspace{2pt}}c@{\hspace{-0.7pt}}c@{\hspace{-1pt}}c@{\hspace{6pt}}}
$Ra$& $Ra/Ra_c$&$N_r$&$l_{max}$& $Rm$ & $Ro_\ell$ 
&$\bar{k}_\phi$&$\bar{k}_s$ & $\bar{k}_z $ 
& $B^2_{peak}$&$B^2_{rms}$
& $|\omega_M/\omega_C|$
&$|\omega_A/\omega_M|$&$u_{\phi,\mbox{sc}}$\\
\multicolumn{14}{c}{$E=6\times10^{-5},Pm=Pr=5$}\\
300   & 10.34  &88  &96  & 67   & 0.002   &4.56&5.03&4.18&0.01&0.52&0.00&\hspace{-0.11in}18.8&  0.00 (0.00)\\
500   & 17.24  &88  &96  & 74   & 0.002   &4.21&4.72&4.53&0.30&1.13&0.03&\hspace{-0.11in}4.27&  0.01 (0.02)\\
800   & 27.58  &88  &96  & 74   & 0.002   &4.61&4.53&4.27&1.8&1.75&0.06&\hspace{-0.11in}2.34&  0.02 (0.07)\\
1000  & 34.48  &128 &120 & 98   & 0.004   &4.14&4.12&4.48&26&2.48&0.21&\hspace{-0.11in}0.67&  0.08 (0.11)\\
2000  & 68.96  &128 &120 & 123  & 0.007   &5.39&3.81&4.79&72&2.95&0.39&\hspace{-0.11in}0.53&  0.14 (0.11)\\
3000  & 103.45 &160 &160 & 169  & 0.009   &6.21&3.97&4.72&112&3.26&0.55&\hspace{-0.11in}0.53&  0.15 (0.11)\\
6000  & 206.90 &160 &160 & 243  & 0.014   &6.49&4.71&4.95&156&3.28&0.74&\hspace{-0.11in}0.58&  0.35 (0.11)\\
8000  & 275.86 &160 &180 & 288  & 0.020   &6.34&4.58&5.96&180&3.29&0.73&\hspace{-0.11in}0.55&  0.41 (0.12)\\
12000 & 413.79 &160 &180 & 365  & 0.024   &6.01&3.51&5.70&222&3.27&0.71&\hspace{-0.11in}0.65&  0.53 (0.15)\\
14000 & 482.76 &160 &180 & 402  & 0.026   &5.87&4.09&5.13&227&3.18&0.76&\hspace{-0.11in}0.75&  0.65 (0.17)\\
16000 & 551.68 &160 &180 & 435  & 0.028   &6.25&4.71&5.42&231&3.12&0.84&\hspace{-0.11in}0.74&  0.72 (0.18)\\
18000 & 620.69 &160 &180 & 456  & 0.032   &6.59&4.53&5.30&234&3.04&0.88&\hspace{-0.11in}0.79&  0.81 (0.19)\\
20000 & 689.66 &160 &180 & 505  & 0.035   &5.97&4.45&5.19&225&2.55&0.77&\hspace{-0.11in}0.91&  0.79 (0.20)\\
21000 & 724.14 &160 &180 & 549  & 0.039   &6.21&4.17&5.24&65&0.62&0.43&\hspace{-0.11in}1.63&  0.19 (0.21)\\
\multicolumn{14}{c}{$E = 1.2 \times 10^{-5},Pm=Pr=1$}\\
300   & 10.34  &90  &96  &78   & 0.004   &4.07&2.89&4.76&0.01&0.31&0.00&\hspace{-0.11in}18.4&  0.00 (0.00)\\
700   & 24.14  &90  &96  &102  & 0.005   &4.71&3.76&5.11&0.69&2.11&0.03&\hspace{-0.11in}3.13&  0.01 (0.02)\\
1000  & 34.48  &132 &144 &112  & 0.006   &5.59&3.87&4.48&55&2.46&0.35&\hspace{-0.11in}0.46&  0.02 (0.03)\\
2500  & 86.21  &168 &160 &174  & 0.011   &5.37&4.14&5.30&90&3.59&0.44&\hspace{-0.11in}0.49&  0.08 (0.07)\\
4000  & 137.93 &180 &168 &224  & 0.017   &6.33&3.84&4.79&135&4.01&0.61&\hspace{-0.11in}0.55&  0.18 (0.11)\\
7000  & 241.38 &192 &180 &312  & 0.026   &6.35&4.04&5.58&194&4.69&0.72&\hspace{-0.11in}0.53&  0.34 (0.24)\\
10000 & 344.83 &192 &180 &384  & 0.033   &6.21&3.09&4.43&248&5.04&0.76&\hspace{-0.11in}0.70&  0.73 (0.32)\\
15000 & 517.24 &192 &180 &500  & 0.045   &7.21&3.79&4.70&260&5.35&0.88&\hspace{-0.11in}0.79&  0.78 (0.37)\\
20000 & 689.66 &192 &180 &573  & 0.052   &7.07&3.13&5.37&258&5.46&0.85&\hspace{-0.11in}0.82&  0.93 (0.45)\\
25000 & 862.07 &192 &180 &655  & 0.061   &7.67&3.03&5.34&255&5.84&0.90&\hspace{-0.11in}0.93&  1.07 (0.51)\\
27000 & 931.03 &192 &180 &698  & 0.065   &8.87&3.51&5.62&223&4.87&0.91&\hspace{-0.11in}0.98&  1.15 (0.50)\\
28000 & 965.52 &192 &180 &775  & 0.073   &8.82&3.39&5.76&89&0.82&0.62&\hspace{-0.11in}1.52&  0.31 (0.52)\\
\hline
\end{tabular}
\caption{Summary of the main input 
		and output parameters used in the dynamo simulations 
		considered in this study. Here, $Ra$
		is the modified Rayleigh number,
		$Ra_c$ is the critical Rayleigh number for 
		the onset of nonmagnetic convection,
		$N_r$ is the number of radial grid points, $l_{max}$ is
		the maximum spherical harmonic degree, $Rm$
		is the magnetic Reynolds number based on the shell thickness and 
		$Ro_l$ is the local Rossby number. The mean 
		$\phi$, $s$, and $z$ wavenumbers are denoted by
		$\bar{k}_\phi$, $\bar{k}_s$ and 
		$\bar{k}_z$ respectively.  
		In addition, $B^2_{peak}$
		is the square of the peak field inside the TC 
		in the saturated dynamo, $B^2_{rms}$ 
		is the measured mean square value of the field in the 
		spherical shell, and $u_{\phi,\mbox{sc}}$ is the
		scaled peak magnitude of the time and azimuthally averaged 
		$\phi$ velocity inside the TC, 
		with its nonmagnetic value given in brackets.}
	\label{parameters}
\end{table}
\endgroup

\begingroup
\setlength{\tabcolsep}{6pt} 
\renewcommand{\arraystretch}{0.7} 
\begin{table}
\centering
\begin{tabular}{llllllllll}
$Ra$ & $\omega_n^2$ & $\omega_C^2$ & $\omega_M^2$ & $-\omega_A^2$ & $\omega_f$ & 
$\omega_s$ & $U_f$ & $U_s$ & $U_{g,z}$\\
 & $(\times10^{10})$ & $(\times10^{8})$ & $(\times10^{8})$ & $(\times10^{8})$ & 
$(\times10^{4})$ & $(\times10^{4})$ & & & \\
\hline
800 & 3.85 & 21.10 & 0.002 & 0.046 & 4.54 & 0 & 7543 & 0  & 7954\\
1000 & 3.21 & 23.27 & 2.00 & 0.51 & 5.16 & 0.331 & 7439 & 293  & 7141, 343\\

12000 & 1.27 & 27.87 & 11.59 & 6.75 & 5.98 & 1.22 & 6810 & 2051  & 8712, 1159\\
18000 & 1.53 & 21.94 & 19.39 & 10.42 & 6.76 & 1.95 & 7729 & 2477 & 8350, 1542\\
21000 & 0.53 & 22.86 & 3.82 & 11.17 & 4.50 & 0 & 7090 & 0 & 8927\\
\hline
\end{tabular}
\caption{Summary of the data for MAC wave measurement in the 
	dynamo models at $E=6\times10^{-5},Pm=Pr=5$. 
	The sampling frequency $\omega_n$ is chosen 
	to ensure that the fast MAC waves are not missed in the measurement 
	of group velocity. The values of $\omega_C^2$, $\omega_M^2$ and 
	$-\omega_A^2$ are calculated using the mean values
of $k_\phi$, $k_s$ and $k_z$. The measured group velocity in the 
$z$ direction ($U_{g,z}$) is compared with the estimated fast ($U_f$) 
or slow ($U_s$) MAC wave velocity.}
\label{cgvalues}
\end{table}
\endgroup

\subsection{Comparisons between the 
dynamo and linear magnetoconvection models}\label{comparison}

The TC may be approximated by a rotating
layer in which convection takes place under an
axial ($z$) magnetic
field with gravity pointing in the downward $z$ direction.
Therefore, a simplified
linear model that studies the 
the evolution of a buoyancy
disturbance in an unstably stratified 
rotating fluid under an axial magnetic field (figure \ref{setup})
provides an insight into the role of wave motions
in TC convection. The notable points of comparison
between the linear and dynamo models are as follows:

\begin{enumerate}
\item \emph{Onset of slow MAC waves}:
The slow waves are detected within the TC
when $|\omega_M/\omega_C| \sim 0.1$. This state
is characterized by the approximate equality
between the peak intensities of slow and fast waves
(figure \ref{onset} (b) and table \ref{comp}),
in good agreement with the linear model where the
two intensities match
 at $|\omega_M/\omega_C| \approx 0.2$ (figure
\ref{linplots} (c)).
Furthermore, both models indicate that
 the parity between the wave motions
persists for higher $\omega_M/\omega_C$.

\item \emph{Suppression of slow MAC waves under strong forcing}:
The slow wave velocity goes to zero within the TC when
$|\omega_A/\omega_M| \approx 1$ (table \ref{comp}), in
agreement with the linear model (figure \ref{linplots} (d)).

\item \emph{Intensity of polar vortices}:
For $|\omega_M/\omega_C| \sim 0.1$, the time and azimuthally
averaged intensity of the polar vortex is much higher than that
in nonmagnetic convection. However, when 
$|\omega_A/\omega_M| \approx 1$,
the vortex intensity decreases appreciably to a value comparable
to that in nonmagnetic convection (see figure \ref{pvavg}
and table \ref{parameters}), in agreement with the behaviour
of the toroidal kinetic energy 
in the linear model (figure \ref{uphi}). 

\end{enumerate}

\begingroup
\setlength{\tabcolsep}{4pt} 
\renewcommand{\arraystretch}{0.8} 
\begin{table}
	\centering 
	\begin{tabular}{c@{\hspace{6pt}}c@{\hspace{6pt}}c@{\hspace{6pt}}c@{\hspace{6pt}}|c@{\hspace{6pt}}c@{\hspace{6pt}}c@{\hspace{6pt}}c@{\hspace{6pt}}}
		\multicolumn{4}{c|}{$E=6\times10^{-5},Pm=Pr=5$}&\multicolumn{4}{c}{$E=1.2\times10^{-5},Pm=Pr=1$} \\
		\hline
		$Ra$   &$|\omega_M/\omega_C|$&$|\omega_A/\omega_M|$
		& $u_{z_s}^\star/u_{z_f}^\star$ 
		&$Ra$   
		& $|\omega_M/\omega_C|$&$|\omega_A/\omega_M|$
		&$u_{z_s}^\star/u_{z_f}^\star$ \\
		1000&	0.21&0.67	 &  1.12  &1000& 0.35&0.46&  1.23   \\
		2000&	0.39& 0.53    &  1.18  &2500& 0.44&0.49&  1.31   \\
		3000&	0.55& 0.53    &  1.27  &4000& 0.61&0.55&  1.28  \\
		
		18000 &0.88&	0.79&    1.47  &25000&0.90&0.93 &  1.44 \\
		20000 &0.77&	0.91&    1.41  &27000&0.91&0.98 &  1.46 \\
		21000 &0.43&	1.63&    0     &28000&0.62& 1.52&  0       \\
		\hline
	\end{tabular}  
	\caption{Ratio of the peak $z$ velocities
		of the slow and fast MAC waves, $u_{z,s}^\star/u_{z,f}^\star$,
		at progressively increasing forcing
		in two dynamo regimes considered in this
		study.} 
	\label{comp}
\end{table}  
\endgroup

\section{Concluding remarks}
\label{concl}

This study investigates convection within the tangent
cylinder in rapidly rotating spherical dynamos through the
analysis of forced MHD waves. Early studies \citep{gafd2006}
had shown that the polar vortices generated in the dynamo
are considerably more intense than that in nonmagnetic convection due
to the formation of isolated off-axis plumes within the TC. It was subsequently
shown that convection would be localized by a laterally varying
$z$ magnetic field \citep{jfm17a},
and the wavenumber at its onset would be
determined by the Ekman number.
The fact that the magnetic field confines convection suggests
localized magnetostrophic balances within the TC.
In this study, it is shown that slow MAC
waves generated at the length scale of
convection support the isolated TC upwellings
in the dipole-dominated
dynamo regime, in turn producing strong anticyclonic polar vortices.
In regions where the magnetic
flux is relatively weak, fast MAC waves are excited, 
although these waves are unable to penetrate
the neutrally buoyant fluid layer that lies above them.

The observed secular variation of the Earth's magnetic
field  \citep{olson1999polar,hulot2002,amit2006} 
suggests maximum drift rates of the polar vortex in the
range 0.6--0.9$^\circ$ yr$^{-1}$. 
The observed peak values are reached in
the present low-inertia dynamo models for strongly driven
convection with $Ra/Ra_c \sim 10^3$ (table \ref{parameters}).
If the forcing is so strong as to cause polarity reversals,
the field within the TC decays away, resulting in much weaker
circulation in the polar regions. 

\section*{Acknowledgements}

This study was supported in part by Research Grant
 MoE-STARS/STARS-1/504 under Scheme for
 Transformational and Advanced Research 
in Sciences awarded by the Ministry of Education (India)
and in part by Research grant CRG/2021/002486 awarded
by the Science and Engineering Research Board (India).
The computations were performed on \emph{SahasraT} and
\emph{Param Pravega}, the supercomputers at the
Indian Institute of Science, Bangalore.

\appendix
\section{Equations for nonmagnetic convection}
\label{nmeqns}
For $Pm=Pr$, the convection-driven dynamo given by equations
\eqref{momentum}--\eqref{div} is compared with
nonmagnetic convection given by the equations
\begin{align}
	E Pr^{-1}  \Bigl(\frac{\partial {\bm u}}{\partial t} + 
	(\nabla \times {\bm u}) \times {\bm u}
	\Bigr)+  {\hat{\bm{z}}} \times {\bm u} = - \nabla p^\star +
	Ra \, T \, {\bm r} \, + E\nabla^2 {\bm u}, \label{mom1} \\
	\frac{\partial T}{\partial t} +({\bm u} \cdot \nabla) T =  \nabla^2 T,  \label{h1}\\
	\nabla \cdot {\bm u} = 0,  \label{div1}
\end{align}
where lengths are scaled by the thickness of the spherical shell $L$, 
time is scaled by $L^2/\kappa$, the velocity $\bm{u}$ is scaled by $\kappa/L$
and $p^\star= \frac1{2} E Pr^{-1} |\bm{u}|^2$.

For a magnetic (dynamo) calculation with the parameters
 $E=6 \times 10^{-5}$, $Pm=Pr=5$, $Ra=3000$, 
 the equivalent non-magnetic calculation
has the parameters $E=6 \times 10^{-5}$, $Pr=5$, $Ra=3000$.

\clearpage
\bibliographystyle{elsarticle-harv} 
%
	\bibliography{tc3}

\end{document}